%% file: main.tex
 \documentclass[conference]{IEEEtran}
\ifCLASSINFOpdf
\else
\fi
\usepackage{algorithmic}

\usepackage{booktabs}
\usepackage{graphicx}
\usepackage[dvips]{psfrag}
\usepackage{pifont}
\usepackage{latexsym}
\usepackage{subfigure}
\usepackage{cite}
\usepackage{psfrag}
\usepackage{url}
\usepackage{stfloats}
\usepackage{amstext,epsfig,amssymb,amsbsy,amsmath}
\usepackage{bbm}
\usepackage{mathrsfs}
\usepackage{array}
\usepackage{eurosym}
\usepackage{pifont}
\usepackage{rotating}
\usepackage{multirow}
\usepackage{multicol}
\usepackage{color}
\usepackage{lscape}
\usepackage{longtable}
\usepackage[para]{threeparttable}
\usepackage{verbatim}
\usepackage{romannum}

\usepackage{tikz}
\usetikzlibrary{plotmarks}
\usetikzlibrary{calc}
\usetikzlibrary{arrows}
\usetikzlibrary{arrows,intersections}
\usetikzlibrary{arrows.meta}
\usetikzlibrary{shapes,decorations}
\usetikzlibrary{positioning}
\usepackage{pgfplots}
\tikzset{sin v source/.style={
  circle,
  draw,
  append after command={
    \pgfextra{
    \draw
      ($(\tikzlastnode.center)!0.5!(\tikzlastnode.west)$)
       arc[start angle=180,end angle=0,radius=0.425ex]
      (\tikzlastnode.center)
       arc[start angle=180,end angle=360,radius=0.425ex]
      ($(\tikzlastnode.center)!0.5!(\tikzlastnode.east)$)
    ;
    }
  },
  scale=1.5,
 }
}

\usepackage{algorithm}

\begin{document}
%
\title{Chance-Constrained AC Optimal Power Flow Integrating HVDC Lines and Controllability}

\author{\IEEEauthorblockN{Andreas Venzke\IEEEauthorrefmark{1}, Lejla Halilba\v{s}i\'{c}\IEEEauthorrefmark{1}, Ad\'{e}lie Barr\'{e}\IEEEauthorrefmark{1}, Line Roald\IEEEauthorrefmark{2} and Spyros Chatzivasileiadis\IEEEauthorrefmark{1}}
 \IEEEauthorblockA{\IEEEauthorrefmark{1}Center for Electric Power and Energy, Technical University of Denmark, Kgs. Lyngby, Denmark\\
\IEEEauthorrefmark{2}Department of Electrical and Computer Engineering, University of Wisconsin-Madison, Madison, USA \\ 
Email: 
\{andven, lhal\}@elektro.dtu.dk, s170075@student.dtu.dk, roald@wisc.edu, spchatz@elektro.dtu.dk}
}

%


\newcommand*{\red}{\textcolor{black}}
\newcommand*{\blue}{\textcolor{black}}
\newcommand*{\bl}{\textcolor{black}}

\maketitle


\begin{abstract}
The integration of large-scale renewable generation has major implications on the operation of power systems, two of which we address in this work. First, system operators have to deal with higher degrees of uncertainty due to forecast errors and variability in renewable energy production. Second, with abundant potential of renewable generation in remote locations, there is an increasing interest in the use of High Voltage Direct Current lines (HVDC) to increase transmission capacity. These HVDC transmission lines and the flexibility and controllability they offer must be incorporated effectively and safely into the system. In this work, we introduce an optimization tool that addresses both challenges by incorporating the full AC power flow equations, chance constraints to address the uncertainty of renewable infeed, modelling of point-to-point HVDC lines, and optimized corrective control policies to model the generator and HVDC response to uncertainty. The main contributions are twofold. First, we introduce a HVDC line model and the corresponding HVDC participation factors in a chance-constrained AC-OPF framework. Second, we modify an existing algorithm for solving the chance-constrained AC-OPF to allow for optimization of the generation and HVDC participation factors. Using realistic wind forecast data, for 10 and IEEE 39 bus systems with HVDC lines and wind farms, we show that our proposed OPF formulation achieves good in- and out-of-sample performance whereas not considering uncertainty leads to high constraint violation probabilities. In addition, we find that optimizing the participation factors reduces the cost of uncertainty significantly.
\end{abstract}

%
\IEEEpeerreviewmaketitle

\section{Introduction}
\subsection{\bl{Motivation}}
Power system operators have to deal with higher degrees of uncertainty. Increasing shares of unpredictable renewable generation, and stochastic loads, can lead to additional costs and jeopardize system security if uncertainty is not explicitly considered and addressed. In addition, with abundant renewable potential being available further away from load centers, e.g. off-shore, High-Voltage Direct Current lines (HVDC) become the preferred technology for transmitting large amounts of renewable energy over longer distances. 
In order to deal with uncertainty, operators carry out both preventive and corrective control actions in their system \cite{Karangelos2016}. HVDC lines and grids can offer corrective control actions in the form of real-time control of active and reactive power flows. The AC optimal power flow (AC-OPF) problem is a key tool for addressing these challenges~\cite{Panciatici2014}. The \mbox{AC-OPF} problem minimizes an objective function (e.g., generation cost) subject to the power system  operational constraints (e.g. limits on the transmission line flows and bus voltages). The goal of this paper is to propose an AC optimal power flow (AC-OPF) formulation that a) considers uncertainty in wind power infeed, b) incorporates an HVDC line model and c) allows for an optimization of the generator and HVDC control response to fluctuations in renewable generation.

%
\subsection{\bl{Literature Review}}
Existing literature considers uncertainty within the OPF problem using methods such as scenario-based or chance-constrained stochastic programming (e.g. \cite{bienstock2014chance, roald2013analytical,schmidli2016stochastic}), robust optimization methods (e.g. \cite{warrington2013,Jabr2015, Lubin2016,lorca2016multistage,louca2018robust,bai2016robust}), \bl{ or distributionally robust optimization (e.g. \cite{lubin2015robust, roald15, zhang2017distributionally, duan2018distributionally})}. Stochastic formulations can include a set of scenarios describing possible realizations of uncertainty, or chance constraints which define a maximum allowable probability of constraint violation. Robust optimization methods on the other hand often assume a pre-defined uncertainty set and secure the system against the worst-case realization inside this set. To deal with the higher complexity arising from the uncertain \bl{parameters}, existing approaches either assume a DC-OPF or use different techniques to achieve a tractable formulation of the AC-OPF under uncertainty. \bl{Examples of approaches utilizing robust optimization for the AC-OPF under uncertainty include \cite{bai2016robust}, which uses explicit maximization to approximate the AC-OPF under uncertainty as a mixed-integer program, and \cite{louca2018robust}, which develops a convex inner approximation by assuming controllable loads at all buses.} 

In this paper we focus on the chance-constrained OPF. Chance-constrained DC-OPF results to a faster and more scalable algorithm, but the DC-OPF is an approximation that neglects losses, reactive power, and voltage constraints. Refs.  \cite{bienstock2014chance} and \cite{roald2013analytical} formulate a chance constrained DC-OPF assuming a Gaussian distribution of the forecast errors. \bl{In \cite{vrakopoulou2013probabilistic}, a combination of randomized and robust optimization is used to achieve a tractable formulation of the chance constrained DC-OPF including N-1 security constraints.} 
\bl{The same authors extended their work to consider a convex relaxation of the non-linear AC OPF problem in \cite{vrakopoulou2013probabilistic2}. Related to convex relaxation ideas in \cite{vrakopoulou2013probabilistic2}, several works \cite{venzke_etal_TPWRS2017, venzke2018HVDC, rostampour2019distributed} have investigated using convex relaxations of the non-convex AC-OPF to achieve a tractable formulation of the chance-constrained AC-OPF. In \cite{venzke_etal_TPWRS2017}, the semidefinite relaxation is applied and both sample-based and analytical solution approaches are discussed. The latter approach is extended to include interconnected AC and HVDC grids in \cite{venzke2018HVDC}. }
\bl{Using the semidefinite relaxation, the work in \cite{rostampour2019distributed} proposes a distributed solving approach for a suitable approximation of the chance constrained AC-OPF using the alternating direction method of multipliers (ADMM).}

\bl{A variety of other approaches have been proposed, typically based either on full or partial linearization \cite{zhang2011chance, duan2018distributionally, schmidli2016stochastic,baker2016distribution,lubin2018chance}. The work in \cite{zhang2011chance} uses linearization and back-mapping to achieve a tractable formulation, while \cite{baker2016distribution,lubin2018chance} obtains analytical reformulations based on linearized AC power flow equations. The work in \cite{duan2018distributionally} uses the Wasserstein metric as distance measure between probability distributions and proposes a tractable formulation of the chance constrained AC-OPF assuming that the true probability distribution is within a defined Wasserstein distance to the empirical distribution based on data samples. The work presented in this paper is most closely related to the approach in \cite{schmidli2016stochastic}, where a linearization used to model the impact of uncertainty is combined with the full AC power flow equations for the forecasted operating point. The papers devises a scalable, iterative solution algorithm, which is observed to produce close to optimal solutions in \cite{roald_andersson_TPWRS2017}.
}  

\subsection{\bl{Contributions}}
Previous work included simultaneous consideration of power injection uncertainty and operation of HVDC in a single optimization problem. For example, Refs.~\cite{roald2017corrective, vrakopoulou_spchatz_isgt2013, wiget_vrakopoulou_pscc2014} consider stochastic OPF formulations which also incorporate HVDC lines and HVDC grids. However, they all assume a DC-OPF formulation. The focus of this paper is to avoid most of these simplifications to the extent that it is possible, and instead use the full non-linear AC power flow equations as the DC-OPF can lead to substantial errors \cite{dvijotham_molzahn_CDC2016}. \blue{The AC-OPF formulation further allows to fully utilize the control capabilities of the HVDC converters, including voltage and reactive power control.} \bl{Our work differs in two important aspects from the work in \cite{venzke2018HVDC}. First, by relying on a non-convex AC-OPF formulation, and a linearization around the forecasted operating point, our approach is more scalable than the semidefinite relaxation used in \cite{venzke2018HVDC}. Second, the work in \cite{venzke2018HVDC} uses a sample-based approach and robust optimization to approximate the chance constraints which can result to conservative results and very low empirical chance constraint violation probabilities. Here, we assume a normal distribution of the forecast errors which can lead to less conservative solutions compliant with the maximum allowable violation probabilities.} In this paper, we propose an iterative chance-constrained AC-OPF for AC grids with HVDC lines, developing further the work described in \cite{schmidli2016stochastic} and elaborated in \cite{roald_andersson_TPWRS2017}. \blue{The main contributions of our work are:
\begin{enumerate}
    \item We integrate an HVDC line model and HVDC corrective control policies in a non-convex chance-constrained AC-OPF framework considering uncertainty in wind power.
    \item We enable optimization of both generator and HVDC participation factors to react to forecast errors within a computationally efficient iterative solution algorithm. 
    \item To improve computational tractability, we propose to utilize a constraint generation method. 
    \item Using realistic wind forecast data and a Monte Carlo Analysis, for 10 and 39 bus systems with HVDC lines and wind farms, we show that (i) not considering uncertainty leads to high constraint violation probabilities whereas our proposed approach achieves compliance with the target chance constraint violation probabilities and (ii) optimizing both generator and HVDC participation factors reduces the cost of uncertainty significantly.
\end{enumerate}}
The structure of this paper is as follows. Section~\ref{SII} states the chance-constrained AC-OPF formulation. In Section~\ref{SIII}, the HVDC line model and HVDC corrective control policy is explained. Section~\ref{SIV} introduces the iterative solution algorithm. Section~\ref{SV} evaluates the performance of the proposed approach on 10 and 39 bus test cases. Section~\ref{SVI} concludes.  

\section[AC OPF with Approx. Chance-Constraints]{Optimal Power Flow Formulation}
\label{SII}
\label{Sec:ACOPF}

This section states the chance-constrained AC-OPF and presents a tractable reformulation of the chance constraints, which is based on the work from \cite{schmidli2016stochastic} and \cite{roald_andersson_TPWRS2017}. For ease of reference, we follow the notation of \cite{roald_andersson_TPWRS2017} wherever possible.  

\subsection{Chance-Constrained AC Optimal Power Flow}
\blue{A power network consists of the set $\mathcal{N}$ of buses, a subset of those denoted by $\mathcal{G}$ have a generator connected. The buses are connected by a set $(i,j) \in \mathcal{L}$ of transmission lines from bus $i$ to $j$. The \mbox{AC-OPF} problem minimizes an objective function (e.g., generation cost) subject to the power system  operational constraints (e.g. limits on the transmission line flows and bus voltages). For a comprehensive review of the AC-OPF problem, the reader is referred to \cite{molzahn_hiskens-fnt2018}.}

The chance-constrained AC-OPF aims at determining the least-cost operating point, which reduces the probability of violating the limits of system components to an acceptable level $\epsilon$ for a range of uncertainty realizations (e.g. $\epsilon = 1\%$). Consequently, the AC-OPF variables, commonly defined in the space of $\mathbf{x}:=\{\mathbf{P},\mathbf{Q},\mathbf{V},\mathbf{\theta}\}$ variables, are not only subject to one possible set of realizations of the uncertain parameters but to a range of uncertain realizations depending on their forecast errors $\mathbf{\omega}$. $\mathbf{P}$, $\mathbf{Q}$, $\mathbf{V}$ and $\mathbf{\theta}$ denote vectors of nodal active and reactive power injections as well as nodal voltage magnitudes and angles, respectively. We assume wind power forecast errors $\mathbf{\omega}$ to be the the only source of uncertainty and to follow a multivariate Gaussian distribution with zero mean and known covariance, as the authors in \cite{roald_andersson_TPWRS2017} have shown is reasonably accurate, even when $\omega$ is not normally distributed. The actual wind power realization $\mathbf{\tilde{P}_{W}}$ is modelled as the sum of its expected value $\mathbf{P_{W}}$ and the forecast error $\mathbf{\omega}$, 
\begin{gather}
    \tilde{P}_{W,i} = P_{W,i} + \omega_i, \quad \forall i \in \mathcal{W}, \label{eq:WIND}
\end{gather}
where $\mathcal{W}$ denotes the subset of network nodes with wind generators connected to them. \blue{Note that our framework readily extends to consider other sources of uncertainty in power injections, e.g. of loads.} We assume that wind power plants are operated with a constant power factor, which means that their reactive power output follows their active power output, i.e., $\tilde{Q}_{W,i} = \gamma(P_{W,i}+\omega_i)$, where the power ratio $\gamma = \sqrt{\frac{1-\cos^{2}\phi}{\cos^{2}\phi}}$ depends on the power factor $\cos{\phi}$ and can be a parameter or an optimization variable. The actual realizations of the OPF decision variables are modelled as the sum of their optimal set-points at the forecasted wind infeed $\mathbf{x}$ and their reactions to a change in wind power injection $\mathbf{\Delta x(\omega)}$, i.e., $\mathbf{\tilde{x}(\omega)} = \mathbf{x} + \mathbf{\Delta x(\omega)}$. This gives rise to the following formulation of the chance-constrained AC-OPF:  
\begin{subequations}
\begin{alignat}{2}
& \min_{\mathbf{x}} \quad  \mathbf{c_{2}^{T}P_{G}^2} + \mathbf{c_{1}^{T}P_{G}} + \mathbf{c_{0}} && \label{obj1} \\ 
& \,  \textrm{s.t.} \quad  f_i(\mathbf{x}) = 0,  && \forall i \in \mathcal{N} \label{eq:CCACOPF_EQ}     \\
& \quad \mathbb{P} (P_{G,k} + \Delta P_{G,k}(\omega) \leq P_{G,k}^{\text{max}} ) \geq 1-\epsilon, \, \,&& \forall k \in \mathcal{G} \label{eq:CCACOPF_INEQ_start} \\
& \quad \mathbb{P} (P_{G,k}^{\text{min}} \leq P_{G,k} + \Delta P_{G,k}(\omega) ) \geq 1-\epsilon, \, \, && \forall k \in \mathcal{G} \label{eq:CCACOPF_INEQ_P2}\\
& \quad \mathbb{P} (Q_{G,k} + \Delta Q_{G,k}(\omega) \leq Q_{G,k}^{\text{max}} ) \geq 1-\epsilon, \, \, && \forall k \in \mathcal{G} \label{eq:CCACOPF_INEQ_Q1}\\
& \quad \mathbb{P} (Q_{G,k}^{\text{min}} \leq Q_{G,k} + \Delta Q_{G,k}(\omega) ) \geq 1-\epsilon, \, \, && \forall k \in \mathcal{G} \label{eq:CCACOPF_INEQ_Q2}\\
& \quad \mathbb{P} (V_{i} + \Delta V_{i}(\omega) \leq V_{i}^{\text{max}} ) \geq 1-\epsilon, \, \, && \forall i \in \mathcal{N} \label{eq:CCACOPF_INEQ_V1}\\
& \quad \mathbb{P} (V_{i}^{\text{min}} \leq V_{i} + \Delta V_{i}(\omega) ) \geq 1-\epsilon, \, \, && \forall i \in \mathcal{N} \label{eq:CCACOPF_INEQ_V2}\\
& \quad \mathbb{P} (P_{L,{ij}} + \Delta P_{L,{ij}}(\omega) \leq P_{L,{ij}}^{\text{max}} ) \geq 1-\epsilon, \, \,  && \forall (i,j) \in \mathcal{L} \label{eq:CCACOPF_INEQ_PL1}\\
& \quad \mathbb{P} (P_{L,{ij}}^{\text{min}} \leq P_{L,{ij}} + \Delta P_{L,{ij}}(\omega)) \geq 1-\epsilon,\, \, \,&& \forall (i,j) \in \mathcal{L} \label{eq:CCACOPF_INEQ_end}
\end{alignat}
\end{subequations}
The chance-constrained AC-OPF \eqref{obj1} -- \eqref{eq:CCACOPF_INEQ_end} minimizes the total generation cost for the forecasted operating point. The terms $\mathbf{P_{G}}$, $\mathbf{Q_{G}}$  denotes the active and reactive power dispatch of the generators, and $\mathbf{c_{2}}$, $\mathbf{c_{1}}$, $\mathbf{c_{0}}$ denote the quadratic, linear and constant cost factors, respectively. The term $\mathbf{P_{L}}$ denotes the active power line flow. Constraint \eqref{eq:CCACOPF_EQ} enforces the $\textrm{n}=2\lvert\mathcal{N}\rvert$ nodal active and reactive power balance equations for the forecasted operating point where $\mathcal{N}$ represents the set of network nodes. Note that we do not explicitly enforce the power balance for $\omega \neq 0$.  Instead, as will be outlined in the following, our formulation ensures satisfaction of the linearized AC equations around the operating point, which in combination with the chosen control policies has been shown to perform well on the non-linear system for reasonable levels of uncertainty \cite{roald_andersson_TPWRS2017}. 
The inequality constraints in \eqref{eq:CCACOPF_INEQ_start} -- \eqref{eq:CCACOPF_INEQ_end} include upper and lower limits on  active and reactive power generation, voltage magnitudes, as well as active power flows $\mathbf{P_L}$. They are formulated as individual chance constraints and enforced with a confidence level of $(1-\epsilon)$. The chance constraints account for the entire range of $\omega$, as they can be  analytically reformulated to tractable deterministic constraints using a first order Taylor expansion, which will be discussed in detail in Section \ref{sec:CC}.
\subsubsection{Affine Policies}
We model the control policies as affine functions of the uncertainty $\omega$. Conventional generators are assumed to balance fluctuations in active power generation according to their generator participation factors $\alpha$ for each generator $k \in \mathcal{G}$ according to
\begin{align}
\tilde{P}_{G,k}(\omega) & = P_{G,k} + \Delta P_{G,k}(\mathbf{\omega})  = P_{G,k} - \alpha_{k}\mathbf{1}\omega + \delta_{k}^{P}, \label{eq:RESERVES}
\end{align}
where the term $\delta^{P}$ denotes the contribution to the compensation of the unknown changes in active power losses, $\mathbf{1}$ represents an all-ones row vector of size $\lvert \mathcal{W} \rvert$. \blue{This generator response mimics Automatic  Generation Control (AGC) commonly used in power system operation.} The generator participation factors $\alpha$ are thus defined w.r.t. to the total wind deviation $\Omega = \sum_{i \in \mathcal{W}} \omega_i$ and can be either pre-determined (e.g., as a result of a reserve procurement) or optimized within the OPF. The condition $\sum_{i \in \mathcal{G}} \alpha_i = 1$ ensures balance of the total power mismatch, i.e., $\sum_{i \in \mathcal{G}}\alpha_i \sum_{i \in \mathcal{W}}\omega_i = \Omega$. Active power losses vary non-linearly with the wind power deviation and are usually compensated by the generator at the reference bus; this results in the loss term $\delta^{P}$ being equal to zero for generators at PV and PQ buses. All other variables of interest $\mathbf{\Delta x(\omega)} := \{ \mathbf{\Delta Q_G},\mathbf{\Delta V},\mathbf{\Delta \theta},\mathbf{\Delta P_{\text{line}}} \}$ are modeled similarly, 
\begin{gather}
    \tilde{x}_i(\omega) = x_i + \mathbf{\Gamma}^{x_i}\omega, \label{eq:AFFINE_POLICY} 
\end{gather}
where $\mathbf{\Gamma}^{x_i}$ is a $(1 \times \lvert \mathcal{W} \rvert)$ vector defining the response of variable $x_i$ to each wind power deviation. In general, the response is modeled as follows: $\mathbf{\Delta x(\omega) = \frac{\partial x}{\mathbf{\partial \omega}} \omega = \Gamma^{x}\omega}$, where $\mathbf{\Gamma^{x}}$ represents a matrix of linear sensitivities w.r.t. $\omega$. The term $\mathbf{\Gamma^{x}}$ also includes expressions for the unknown changes in active power losses $\delta^{P}$ and is derived from the first order Taylor expansion of the AC power flow equations around the forecasted operating point,
\begin{gather}
\begin{bmatrix}
\mathbf{\Delta P} \\
\mathbf{\Delta Q} \\
\end{bmatrix} = 
\mathbf{J} \Big\rvert_{\mathbf{x^{*}}}
\begin{bmatrix}
\mathbf{\Delta\theta} \\
\mathbf{\Delta V} \\
\end{bmatrix}. \label{eq:TAYLOR_ACPF}
\end{gather}
The term $\mathbf{J}$ denotes the Jacobian matrix. The left-hand side of \eqref{eq:TAYLOR_ACPF} can also be expressed in terms of the wind deviation $\omega$, the power ratio $\gamma$, the generator participation factors $\alpha$ as well as the unknown nonlinear changes in active and reactive power (i.e., $\delta^{P}$, $\mathbf{\Delta Q}$),
\begin{gather}
\begin{bmatrix}
\mathbf{I} \\
\mathbf{diag(\gamma)} 
\end{bmatrix} \mathbf{\omega} +
\begin{bmatrix}
\mathbf{-\alpha H} \\
\mathbf{0} \\
\end{bmatrix} \mathbf{\omega} + 
\begin{bmatrix}
\mathbf{\delta^{P}} \\
\mathbf{\Delta Q} \\
\end{bmatrix} = 
\mathbf{\Psi} \omega +
\begin{bmatrix}
\mathbf{\delta^{P}} \\
\mathbf{\Delta Q} \\
\end{bmatrix}. \label{eq:LFTHAND_SIDE}
\end{gather}
The terms $\mathbf{I}$, $\mathbf{H}$ and $\mathbf{0}$ denote $(\lvert\mathcal{N}\rvert\times\lvert\mathcal{W}\rvert)$ identity, all-ones and zero matrices, respectively. The matrix of Generation Distribution Factors (GDF) $\mathbf{\Psi}$ depends linearly on $\alpha$ and $\gamma$ (for a detailed derivation refer to \cite{Line_PhD}). \bl{Replacing the left-hand side in \eqref{eq:TAYLOR_ACPF} with \eqref{eq:LFTHAND_SIDE} yields:
\begin{gather}
\mathbf{\Psi} \omega +
\begin{bmatrix}
\mathbf{\delta^{P}} \\
\mathbf{\Delta Q} \\
\end{bmatrix} = 
\mathbf{J} \Big\rvert_{\mathbf{x^{*}}}
\begin{bmatrix}
\mathbf{\Delta\theta} \\
\mathbf{\Delta V} \\
\end{bmatrix} \label{eq:NEW}
\end{gather}}
In accordance with common practices in power system operations, some variables are assumed not to change under different wind power realizations, such as the voltage magnitude at PV and reference buses, the voltage angle at the reference bus and the reactive power injection at PQ buses. We summarize the nonzero changes of unknown active and reactive power injections in $\mathbf{\delta} := [\delta^{P}_{ref} \; \Delta Q_{ref}\; \mathbf{\Delta Q_{PV}^{^T}]^{T}}$. Analogously, $\mathbf{\Delta \hat{x}}$ denotes the nonzero changes in voltage magnitudes and angles, i.e., $\mathbf{\Delta \hat{x}} := [\mathbf{\Delta \theta_{PV}^{T}} \; \mathbf{\Delta \theta_{PQ}^{T}} \; \mathbf{\Delta V_{PQ}^{T}}]^{\mathbf{T}}$.  Rearranging the resulting system of equations in \eqref{eq:NEW} according to the groups of zero and nonzero elements
\begin{gather}
\begin{bmatrix}
\mathbf{\delta} \\
\mathbf{0}
\end{bmatrix} =
\begin{bmatrix}
\mathbf{J_{mod}^{\Romannum{1}}} & \mathbf{J_{mod}^{\Romannum{2}}} \\
\mathbf{J_{mod}^{\Romannum{3}}} & \mathbf{J_{mod}^{\Romannum{4}}}
\end{bmatrix}
\begin{bmatrix}
\mathbf{0}  \\
\mathbf{\Delta \hat{x}} 
\end{bmatrix} -
\begin{bmatrix}
\mathbf{\Psi_{mod}^{\Romannum{1}}} \\
\mathbf{\Psi_{mod}^{\Romannum{2}}}
\end{bmatrix} \mathbf{\omega}, \label{eq:TAYLOR_ACPF_REARR} 
\end{gather}
allows us to derive expressions \eqref{eq:Sensitivity1} and \eqref{eq:Sensitivity2} for the change in variables as a function of the uncertainty $\omega$.
\begin{subequations}
\begin{gather}
\mathbf{\Delta \hat{x}} = \Big( \mathbf{J_{mod}^{\Romannum{4}}}\Big)^{-1} \mathbf{\Psi_{mod}^{\Romannum{2}}} \mathbf{\omega} = \mathbf{\Gamma^{\hat{x}}} \mathbf{\omega} \label{eq:Sensitivity1} \\
\mathbf{\delta} =  \Big( \mathbf{J_{mod}^{\Romannum{2}}} (\mathbf{J_{mod}^{\Romannum{4}}})^{-1}
\mathbf{\Psi_{mod}^{\Romannum{2}}} - \mathbf{\Psi_{mod}^{\Romannum{1}}} \Big) \mathbf{\omega} = \mathbf{\Gamma^{\delta}} \mathbf{\omega} \label{eq:Sensitivity2} 
\end{gather}
\end{subequations}
$\mathbf{J_{mod}}$ and $\mathbf{\Psi_{mod}}$ denote the modified Jacobian and GDF matrices, where the rows and columns have been rearranged according to $\delta$ and $\mathbf{\Delta \hat{x}}$. Thus, the linear sensitivities $\mathbf{\Gamma^{x}}$ depend on the GDF matrix $\mathbf{\Psi}$, which is a linear function of the generator participation factors $\alpha$ and the power ratio $\gamma$.

\subsubsection{Reformulating the Chance Constraints} \label{sec:CC}
Given the linear dependency of the OPF variables on $\omega$ in the region around the operating point and the assumption of a multivariate normal distribution for $\omega$, we are able to reformulate the individual chance constraints \eqref{eq:CCACOPF_INEQ_start}--\eqref{eq:CCACOPF_INEQ_end} to tractable deterministic constraints. The linear chance constraint $\mathbb{P}(x_i + \mathbf{\Gamma}^{x_i}(\mathbf{\Psi}) \omega \leq x_i^{\text{max}}) \geq 1 - \epsilon$ is reformulated to
\begin{gather}
    x_i \leq x_i^{\text{max}} - \Phi^{-1}(1-\epsilon)\sqrt{\mathbf{\Gamma}^{x_i} \mathbf{\Sigma} (\mathbf{\Gamma}^{x_{i}})^{\mathbf{T}}}, \label{eq:CC_REFORMULATION} 
\end{gather}
where $\Phi^{-1}$ denotes the inverse cumulative distribution function of the Gaussian distribution. It can be observed that the original constraint $x_i \leq \overline{x_i}$ is tightened by an \textit{uncertainty margin} $\lambda^{x_i} := \Phi^{-1}(1-\epsilon)\sqrt{\mathbf{\Gamma}^{x_i} \mathbf{\Sigma} (\mathbf{\Gamma}^{x_{i}})^{\mathbf{T}}}$, which secures the system against variations in wind infeed \cite{schmidli2016stochastic}. Given the dependency of $\mathbf{\Gamma^{x}}$ on $\mathbf{\Psi}$, optimizing over the generation response $\alpha$ explicitly represents its impact on the uncertainty margins of the remaining variables within the optimization.    

\section[HVDC]{HVDC Line Modeling}
\label{SIII}
In this section, we present a model to include HVDC lines in the chance-constrained AC-OPF and we introduce HVDC participation factors to allow for corrective control.
\begin{figure}
\begin{center}
\begin{tikzpicture}[font = \footnotesize,dot/.style={draw,circle,minimum size=1.5mm,inner sep=0pt,outer sep=0pt,fill=white}]
\draw (0,0) coordinate[dot](n1)--++ (1.5,0) coordinate(n2);
\path (n2) --++ (0,-1.0) coordinate(n3);
\draw (n2) --++ (0,-0.5) node[below,sin v source]{};
\draw (n3) --++ (0,-0.3) --++ (-0.2,0) --++ (0.4,0);
\path (n2) --++ (-0.3,0.2) coordinate (nP1);
\path (n2) --++ (-0.3,-0.2) coordinate (nQ1);
\draw[->] (nP1) --++ (-0.5,0) node[above]{$P_{C,i}$};
\draw[->] (nQ1) --++ (-0.5,0) node[below]{$Q_{C,i}$};
\path (0,0) --++ (0,0) node[above]{$i$};
\path (n2) --++ (4,0) coordinate(n4);
\draw (n4) --++ (0,-0.5) node[below,sin v source]{};
\path (n4) --++ (0,-1.0) coordinate(n5);
\draw (n5) --++ (0,-0.3) --++ (-0.2,0) --++ (0.4,0);
\path (n4) --++ (0.3,0.2) coordinate (nP2);
\path (n4) --++ (0.3,-0.2) coordinate (nQ2);
\draw[->] (nP2) --++ (0.5,0) node[above]{$P_{C,j}$};
\draw[->] (nQ2) --++ (0.5,0) node[below]{$Q_{C,j}$};
\draw (n4) --++ (1.5,0) coordinate[dot](n6);
\path (n6) --++ (0,0) node[above]{$j$};
\path (n2) --++ (0.75,0) coordinate(n7);
\path (n7) --++ (1.25,-0.3) node[below]{DC system};
\draw (n7) --++ (2.5,0) --++ (0,-1) --++ (-2.5,0) --++ (0,1);
\path (n7) --++ (1.25,-1.0) coordinate(n9);
\draw[->] (n9) --++ (0,-0.3) node[right]{$P_{\text{loss}}$};
\end{tikzpicture}
  \vspace{-0.4cm}
\end{center}
    \caption{\blue{HVDC line model connecting AC bus $i$ to AC bus $j$ with active HVDC converter injections $\mathbf{P_{C}}$, reactive HVDC converter injections $\mathbf{Q_{C}}$ and an active loss term $P_{\text{loss}}$.}}
    \label{HVDC}
    \vspace{-0.2cm}
\end{figure}
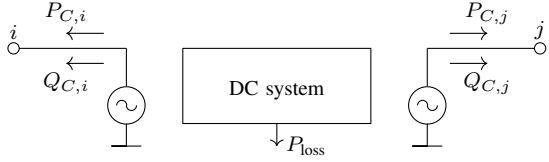 
\blue{We assume that the HVDC lines are modeled as presented in Fig.~\ref{HVDC} with individual active and reactive power injections $\mathbf{P_{C}}$, $\mathbf{Q_{C}}$  at the two AC buses the HVDC line is connected to and a lumped loss term $P_{\text{loss}}$ for the DC system losses. The set $c \in \mathcal{N}_{\text{C}}$ denotes the HVDC converter and for each two HVDC converter comprising an HVDC line the set $(i,j) \in \mathcal{L}_{C}$ denotes the AC buses the HVDC converters are connected to, respectively.} We approximate the active and reactive power capability of the converter as a rectangular box with the following constraints:
\begin{subequations}
\begin{alignat}{2}
   & P_{C,c}^{\text{min}} \leq  P_{C,c} \leq  P_{C,c}^{\text{max}} \quad && \forall c \in \mathcal{C}\\
   &  Q_{C,c}^{\text{min}}  \leq  Q_{C,c} \leq  Q_{C,c}^{\text{max}} \quad && \forall c \in \mathcal{C}
\end{alignat}
\end{subequations}
Expressing the lower and upper active and reactive HVDC converter limits $\mathbf{P_{C}^{\text{min}}}$, $\mathbf{P_{C}^{\text{max}}}$, $\mathbf{Q_{C}^{\text{min}}}$, $\mathbf{Q_{C}^{\text{max}}}$ as a function of the nominal converter rated power $\mathbf{S_{C}^{\text{nom}}}$ and assuming that the lower and upper bounds on active power are symmetric (i.e.~$\mathbf{P_{C}^{\text{min}}} = - \mathbf{P_{C}^{\text{max}}}$) yields:
\begin{subequations}
\begin{alignat}{2}
   -m_{p,c} S_{C,c}^{\text{nom}} \leq & P_{C,c} \leq  m_{p,c} S_{C,c}^{\text{nom}} \quad  && \forall c \in \mathcal{C} \\
   m_{q,c}^{\text{min}} S_{C,c}^{\text{nom}} \leq  & Q_{C,c} \leq  m_{q,c}^{\text{max}} S_{C,c}^{\text{nom}} \quad && \forall c \in \mathcal{C} 
\end{alignat}
\end{subequations}
\begin{figure}
\center
\resizebox{0.37\textwidth}{!}{
\begin{tikzpicture}[
	font = \normalsize,
    thick,
    >=stealth',
    dot/.style = {
      draw,
      fill = gray!40,
      circle,
      inner sep = 2pt,
      minimum size = 0pt
    }
  ]
  \draw[white,fill=blue!40] (-1.28,0.6) -- (-1.28,-0.8) -- (1.28,-0.8) -- (1.28,0.6);
  \draw[white,fill=white] (-2,0.6) -- (+2.5,0.6) -- (4,2) -- (-4,2);
  \draw[white,fill=white] (-2,-0.8) -- (+2.5,-0.8) -- (4,-1.55) -- (-4,-1.55);
  \draw[blue,dashed] (0,0) ellipse (1.5 and 1.5);
  \draw[->] (-2,0) -- (2,0) coordinate[label = {right:$P_{C,c}$}] (xmax);
  \draw[->] (0,-1.8) -- (0,1.8) coordinate[label = {above:$Q_{C,c}$}] (ymax);
  \draw[blue,dashed] (-2,0.6) -- (2,0.6) coordinate[label = {right:$m_{q,c}^{\text{max}} S_{C,c}^{\text{nom}}$}] (ymax);
  \draw[blue,dashed] (-2,-0.8) -- (2,-0.8) coordinate[label = {right:$ m_{q,c}^{\text{min}} S_{C,c}^{\text{nom}} $}] (ymax);
	  \path[blue] (0,0) -- (1.8,1.4)coordinate[label = {}] (ymax);
	  
	    \draw[blue,dashed] (-1.28,1.7) -- (-1.28,-1.7) coordinate[label = {left:$-m_{p,c} S_{C,c}^{\text{nom}} $}] (ymax);
  \draw[blue,dashed] (1.28,1.7) -- (1.28,-1.7)  coordinate[label = {right:$m_{p,c} S_{C,c}^{\text{nom}}  $}] (ymax);
	   \draw[black] plot [smooth] coordinates {(-1.2,0.2) (-1.8,0.4) (-2.0,1.0)};
	 \path[black] (0,0) -- (-2.5,0.8) coordinate[label = {above: $\begin{matrix} \text{Feasible} \\ \text{operating region} \end{matrix}$}] (ymax);
\end{tikzpicture}}
\vspace{-0.2cm}
\caption{Active and reactive power capability curve of HVDC converter $c$ \cite{imhof2015voltage}.}
\vspace{-0.2cm}
\label{PQ_Conv}
\end{figure}
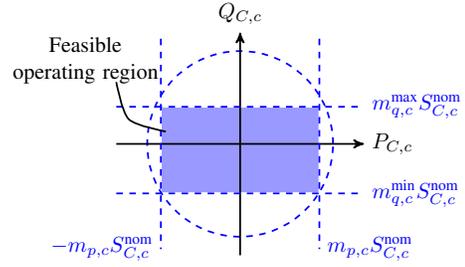
The resulting feasible operating region is visualized in Fig.~\ref{PQ_Conv}. For a more detailed modeling of the active and reactive power capability of HVDC converter the interested reader is referred to \cite{beerten_TPWRS2012}. In order to link the active power injections between the two AC buses that the HVDC line is connected to, an active power balance constraint has to be included. 
To model the DC system losses $P_{\textbf{loss}}$, we use a constant loss term $a$ defined as a share of the nominal apparent power rating for buses $(i,j) \in \mathcal{L}_{C}$ and converters $c \in \mathcal{C}$:
\begin{align}
     P_{\text{HVDC},i}+P_{\text{HVDC},j}+& P_{\text{loss},c}=0 \label{eq:DC_BALANCE_LOSS} \quad  \text{with } P_{\text{loss},c} = 2 a  S_{C,c}^{\text{nom}}
\end{align}
This term gives an estimate of the HVDC converter losses. Note that we neglect the DC line losses. The reactive power injections at both AC buses  $(i,j) \in \mathcal{L}_\text{C}$ can be chosen independently from each other within the HVDC converter limits. To allow for corrective control, we assign a participation factor $\beta_c$ for each HVDC converter $c \in \mathcal{C}$ similarly to the case of generators. As the HVDC line itself cannot generate active power, the participation factor is positive at one end of the HVDC line and negative at the other end, i.e. $\beta_i = -\beta_j$ for buses $(i,j) \in \mathcal{L}_{C}$. This controllability can be used to e.g. reroute power to reduce congestion in case of different forecast error realizations. The GDF matrix $\mathbf{\Psi}$ is modified as follows:
\begin{gather}
\mathbf{\Psi} = 
\begin{bmatrix}
\mathbf{I} - (\alpha + \beta) \mathbf{H} \\
\mathbf{diag(\gamma)}
\end{bmatrix} \label{eq:PSI_MOD}
\end{gather}
The HVDC participation factors $\beta$ are nonzero only for the converter connected AC buses and its sign depends on which end of the HVDC line the AC bus is connected to. Similar to the engineering constraints of the AC grid, the converter limits need to be considered as chance constraints in order to ensure secure operation with sufficient probability throughout the uncertainty range, e.g.,
\begin{subequations}
\begin{alignat}{2}
   & \mathbb{P} ( -m_{p,c} S_{C,c}^{\text{nom}} \leq   P_{C,c}  + \beta_c \omega ) \quad &&\geq 1-\epsilon \quad  \forall c \in \mathcal{C},\\
  &  \mathbb{P} ( m_{p,c} S_{C,c}^{\text{nom}} \geq  P_{C,c} + \beta_c \omega  ) \quad && \geq 1-\epsilon \quad \forall c \in \mathcal{C}.
\end{alignat}
\end{subequations}
These can be reformulated for each converter $c \in \mathcal{C}$:
\begin{subequations}
\begin{alignat}{2}
 -m_{p,c} S_{C,c}^{\text{nom}} & +   \Phi^{-1}(1-\epsilon) \sqrt{\beta_c \mathbf{1} \mathbf{\Sigma} \beta_c \mathbf{1}^T} \leq P_{C,c}, && \label{eq:UM_HVDC1}\\
  m_{p,c} S_{C,c}^{\text{nom}} & -    \Phi^{-1}(1-\epsilon) \sqrt{\beta_c \mathbf{1} \mathbf{\Sigma} \beta_c \mathbf{1}^T} \geq P_{C,c} . && \label{eq:UM_HVDC2} 
\end{alignat}
\end{subequations}
Note that the uncertainty margins $\lambda^{\mathbf{P_{C}}}$ introduced in \eqref{eq:UM_HVDC1} and \eqref{eq:UM_HVDC2} depend linearly on the HVDC participation factor $\beta$. The degree of controllability is determined by $\alpha$ and $\beta$, both of which can be either pre-determined or optimized within the chance-constrained AC-OPF.

\section{Iterative Chance-constrained AC-OPF optimizing Generator and HVDC control policies} 
\label{SIV}
The reformulated chance-constrained AC-OPF \eqref{obj2} -- \eqref{eq:CCACOPF_INEQ3} considering HVDC lines extends the variable set $\mathbf{x}$ to include the active and reactive power set-points of the HVDC converters $[\mathbf{P_{C}}$, $\mathbf{Q_{C}}]$:
\begin{subequations}
\begin{alignat}{2}
\min_{\mathbf{x}} \quad  & \mathbf{c_{2}^{T}P_{G}^2} + \mathbf{c_{1}^{T}P_{G}} + \mathbf{c_{0}} && \label{obj2} \\ 
\textrm{s.t.}  & \mathbf{f^{ac}}(\mathbf{x}) = 0  &&  \label{eq:CCACOPF_EQAC}     \\
& \mathbf{f^{dc}}(\mathbf{P_{C}}) = 0  &&  \label{eq:CCACOPF_EQDC}     \\
& \mathbf{x} \leq \mathbf{x^{\text{max}}} - \mathbf{\lambda^{x}(\alpha,\beta)} \quad && \label{eq:CCACOPF_INEQ2} \\
& \mathbf{x} \geq \mathbf{x^{\text{min}}} + \mathbf{\lambda^{x}(\alpha,\beta)} \quad && \label{eq:CCACOPF_INEQ3} 
\end{alignat}
\end{subequations}
If the corrective control actions provided by conventional generators and HVDC lines are optimized within the same framework, the participation factors $\alpha$ and $\beta$ extend the variable set $\mathbf{x}$ and the following additional equations are included:
\begin{subequations}
\begin{alignat}{2}
& \beta_i = -\beta_j \quad &&  \forall(i,j) \in \mathcal{L}_{C} \label{eq:alpha_beta_1}\\
& \alpha_k \geq 0 && \forall k \in \mathcal{G} \\
& \sum_{k \in \mathcal{G} } \alpha_k = 1 \label{eq:alpha_beta_3}
\end{alignat}
\end{subequations}
The problem \eqref{obj2} -- \eqref{eq:CCACOPF_INEQ3} introduces for each HVDC line an additional power balance equation \eqref{eq:CCACOPF_EQDC} according to \eqref{eq:DC_BALANCE_LOSS} considering the losses in the DC system. All inequality constraints are tightened with their corresponding uncertainty margins $\mathbf{\lambda^{x}(\alpha,\beta)} =  [\mathbf{\lambda^{P_G}}, \, \mathbf{\lambda^{Q_G}}, \,  \mathbf{\lambda^{V}}, \,  \mathbf{\lambda^{P_{L}}}, \,  \mathbf{\lambda^{P_C}}](\alpha,\beta)$. The uncertainty margins do not only depend on the generator and HVDC participation factors but also on the Jacobian matrix of the AC power flow equations as can be observed in \eqref{eq:Sensitivity1} and \eqref{eq:Sensitivity2}. Including the Jacobian terms as optimization variables would introduce even more non-linearities in the AC-OPF and thus, substantially increase the complexity of the problem. To this end, the authors in \cite{roald_andersson_TPWRS2017} have introduced a computationally efficient iterative solution algorithm, which decouples the uncertainty assessment (i.e., the derivation of the uncertainty margins) from the optimization.     

\begin{algorithm}
\caption{Iterative Chance-Constrained AC-OPF Optimizing Generator and HVDC Corrective Control Policies}
\begin{algorithmic}[1]
\STATE Set iteration count: $k \leftarrow 0$  
\STATE initialize $\mathbf{\lambda}^{\mathbf{x},0} = \mathbf{0}$
\WHILE{$\lvert \lvert \mathbf{\lambda}^{\mathbf{x},k}-\mathbf{\lambda}^{\mathbf{x},k-1} \rvert \rvert_{\infty} > \rho$}
\IF{$k=0$}
\STATE solve \eqref{obj2} -- \eqref{eq:CCACOPF_INEQ3} for $\mathbf{x} \setminus \{ \alpha, \beta\}$  and obtain $\mathbf{x}^{0}_{\text{opt}}$
\STATE evaluate Jacobian at $\mathbf{x}^{0}_{\text{opt}}$  
\ELSE
\STATE include $\mathbf{\lambda}^{\mathbf{x},k}(\alpha^{k},\beta^{k})$ according to \eqref{eq:CCACOPF_INEQ2} and \eqref{eq:CCACOPF_INEQ3} 
\STATE solve \eqref{obj2} -- \eqref{eq:CCACOPF_INEQ3}, \eqref{eq:alpha_beta_1} -- \eqref{eq:alpha_beta_3} to obtain $\mathbf{x}^{k}_{\text{opt}}$
\STATE evaluate Jacobian, $\mathbf{\Gamma}^{k}_{\text{opt}}$ and $\mathbf{\lambda}^{k}_{\text{opt}}$ at $\mathbf{x}^{k}_{\text{opt}}$, $\mathbf{\alpha}^{k}_{\text{opt}}$ and $\mathbf{\beta}^{k}_{\text{opt}}$
\ENDIF
\STATE derive expressions for $\mathbf{\Gamma}^{\mathbf{x},k+1}$ and $\mathbf{\lambda}^{\mathbf{x},k+1}$ as functions of optimization variables $\alpha^{k+1}$ and $\beta^{k+1}$ 
\STATE $k \leftarrow k+1$
\ENDWHILE.
\end{algorithmic}
\label{ALG}
\end{algorithm}
To maintain computational efficiency, we extend the iterative framework of \cite{roald_andersson_TPWRS2017} and evaluate the Jacobian at each iteration for the current operating point. In \cite{roald_andersson_TPWRS2017}, the uncertainty margins were constants and were computed in an outer iteration. In the current paper, the sensitivity factors are constants, while  $\alpha$ and $\beta$ are kept as optimization variables, which allows us to optimize these at the expense of adding non-linear (but convex) second order cone (SOC) terms. We define the steps in Algorithm~\ref{ALG}, where subscript \textit{opt} denotes the optimal solution of an OPF. The algorithm converges as the change in uncertainty margins between two consecutive iterations falls below a defined tolerance value $\rho$.  

If we include the participation factors as optimization variables in the iterative solution algorithm, the right hand sides of the inequalities \eqref{eq:CCACOPF_INEQ2}-\eqref{eq:CCACOPF_INEQ3} are a non-linear function of the participation factors in the OPF problem. As a result, the computational complexity is increased. To maintain scalability, we propose to use a constraint generation method to solve the AC-OPF in each step of Algorithm~\ref{ALG} based on \cite{roald2017corrective}: First, we solve the AC-OPF excluding all uncertainty margins (i.e. they are set to zero), except the uncertainty margins for the generators \eqref{eq:CCACOPF_INEQ_start} -- \eqref{eq:CCACOPF_INEQ_P2} and the HVDC active power \eqref{eq:UM_HVDC1} -- \eqref{eq:UM_HVDC2}. Note that for these constraints, we can simplify the uncertainty margin to a linear function in the participation factors and including these is computationally cheap. Then, based on the OPF results, we iteratively evaluate all uncertainty margins for the optimized values of the participation factors. Only those inequality constraints in \eqref{eq:CCACOPF_INEQ2}--\eqref{eq:CCACOPF_INEQ3}, which are violated for the current optimized state variables and participation factors are included in the next OPF problem. The OPF problem is resolved until the solution complies with all constraints \eqref{eq:CCACOPF_INEQ2}--\eqref{eq:CCACOPF_INEQ3}. As we will show in Section~\ref{39bus_results}, this allows us to reduce the number of considered uncertainty margins significantly and maintain scalability of our approach. 
\label{Sol_algo}

\bl{In case the actual true uncertainty distribution cannot be well captured by a Gaussian distribution or only limited forecast data is available, it is possible to formulate distributionally robust versions of the chance constraints. An increasing number of papers consider distributional robustness, including e.g. \cite{lubin2015robust, roald15, zhang2017distributionally, duan2018distributionally}. 
Distributional robustness can be understood in terms of the ambiguity regarding the parameters of the distribution \cite{lubin2015robust, zhang2017distributionally} or regarding the type of distribution \cite{roald15}. Different types of ambiguity and associated uncertainty sets will result in different problem reformulations, which may be more or less tractable. 
The work in \cite{duan2018distributionally} uses the Wasserstein metric as distance measure between probability distributions and proposes a tractable formulation of the chance constrained AC-OPF assuming that the true probability distribution is within a defined Wasserstein distance to the empirical distribution based on data samples. Note that some distributionally robust approaches, such as the one presented in \cite{roald15}, allows for a similar reformulation of the individual chance constraints \eqref{eq:CCACOPF_INEQ_start}--\eqref{eq:CCACOPF_INEQ_end}. 
Essentially, it is possible to obtain valid chance-constraint reformulation for any random variables with finite mean and covariance by replacing $\Phi^{-1}(1-\epsilon)$ by a different (constant) function. This will lead to a more conservative, but safe solution. 
}
\section{Simulations and Results}
\label{SV}
We specify the simulation setup. In the first part, we show the benefit of optimizing the generator participation factors for the proposed iterative chance-constrained AC-OPF for a 10 bus system. In the second part, we include an HVDC line in this system to relieve congestion in the AC system and investigate optimizing both the generator and HVDC control policies and, in addition, the convergence behaviour of the iterative solution algorithm. In the third part, we consider an IEEE 39 bus system and evaluate the benefit of controllability.
 \subsection{Simulation Setup}
 To evaluate the performance of the proposed approaches we use two metrics. First, we compute the cost of uncertainty  which is the increase in generation cost by including chance constraints. \bl{Let $f_0$ denote the objective value of the AC-OPF without considering uncertainty, i.e. all uncertainty margins are set to zero: $\lambda^x = 0$. Note that we will refer to this OPF problem as AC-OPF ($\lambda^x = 0$, w/o uncertainty) in the following. Let $f_U$ denote the objective value of the chance-constrained AC-OPF, i.e. all uncertainty margins are computed according to the presented iterative solution algorithm, and the OPF formulation takes into account the uncertainty in the wind power injections. Then, we can compute the cost of uncertainty as follows:}
 \begin{align}
     \bl{\text{Cost of Uncertainty} = \tfrac{f_U - f_0}{f_0} \times 100 \%}
 \end{align}
 \bl{The cost of uncertainty is expressed in percent and is always larger or equal to zero as the resulting tightening of the right hand side of \eqref{eq:CCACOPF_INEQ2} and \eqref{eq:CCACOPF_INEQ3} shrinks the OPF feasible space.}
Second, we perform an in- and out-of-sample analysis to compute the empirical individual chance constraint violation probability. To determine the mean and variance of the Gaussian distribution, we use a limited amount of samples from realistic wind forecast data. For the in-sample analysis we draw 10'000 samples from this Gaussian distribution and evaluate the performance (i.e. the occurring constraint violations) of our proposed OPF formulation. For the out-of-sample analysis we use 10'000 samples from same database of realistic wind forecast data. This allows a first assessment of how our proposed OPF formulation performs if the wind realizations do not exactly match a Gaussian distribution. For the in- and out-of-sample Monte Carlo Analysis we assume a minimum violation limit of 0.1\% to exclude numerical errors. Note that for each type of individual chance constraint, we report the maximum observed empirical violation probability. To compute the constraint violations, we use AC power flows in MATPOWER \cite{zimmerman2011matpower}. The maximum allowable constraint violation limit is set to $\epsilon = 5\%$. We consider a convergence criterion of $\rho = 10^{-5}$.  All simulations are carried out on a laptop with processor Intel(R) Core(TM) i7-7820HQ CPU 2.90 Ghz and 32GB RAM. The optimization problems are implemented with YALMIP\cite{Lofberg2004} in MATLAB and are solved with IPOPT \cite{Wachter2006}. The wind farm power factor $\gamma$ is set to 1.
 
\begin{figure}[!t]
\center
\resizebox{0.4\textwidth}{!}{
\begin{tikzpicture}[font = \normalsize]
\draw (0,0) coordinate(node6) --++ (0,2.5) coordinate(node5);
\draw (node5) --++ (0,4.0) coordinate(node4);
\draw(node5) --++ (-1,0) node[left]{5} --++ (0.5,0);
\path (node5) --++ (-0.5,0) coordinate(node5g);
\draw (node5g) --++ (0,-0.5) node[below,sin v source]{}; 
\path (node5) --++ (-0.75,-0.75) node[left]{G3};
\draw (node5) --++ (1.0,0) coordinate(L5);
\draw (L5) --++ (0.5,0);
\draw (node6) --++ (-1,0) node[left]{6} --++ (2,0);
\draw (node4) --++ (-1,0) node[left]{4}--++ (3,0);
\draw (node4) --++ (0.75,0) --++ (0,-2);
\path (node4) --++ (1.5,0) coordinate(node4w);
\draw (node4w) --++ (0,-0.5) node[below,sin v source]{};
\path (node4w) --++ (0.25,-0.75) node[right]{W2};
\draw[-{Latex[length=0.25cm]}](node6) --++(0,-0.75)node[below]{};
\draw[-{Latex[length=0.25cm]}](L5) --++(0,-0.75)node[below]{};
\draw (L5) --++ (0,2) coordinate (node3);
\draw (node3) --++ (0.5,0) coordinate (node3g);
\draw (node3g) --++ (0,-0.5) node[below,sin v source]{};
\path (node3g) --++ (0.25,-0.75) node[right]{G2};
\draw (node3g) --++ (1.75,0) coordinate (node3HVDC);
\draw (node3g) --++ (1.0,0) --++ (0,0.5) --++ (5,0) --++ (0,1.5);
\draw (node3HVDC) --++ (-0.25,0) --++ (0,-0.5) --++ (1.75,0) --++ (0,0.5);
\draw[thick]  (node3) --++ (-0.5,0) node[left]{3};
\draw (node6) --++ (0.75,0) --++ (0,-1.0) --++ (2.5,0) --++ (0,-0.5);
\draw (node6) --++ (0.33,0) --++ (0,0.5) --++ (2.2,0) --++ (0,4);
\path (node6) --++ (3,-1.5) coordinate (node7);
\draw (node7) --++ (1.5,0) --++ (0,0.5) --++ (1.5,0) --++ (0,-0.5);
\path (node7) --++ (3,0) coordinate (node8);
\draw (node7) --++ (-0.25,0) node[left]{7} --++ (2.5,0);
\draw (node8) --++ (-0.25,0) node[left]{8} --++ (2.0,0);
\draw (node8) --++ (1,0) --++ (0,1.0);
\path (node8) --++ (0.5,3.0)coordinate(node9);
\draw (node9) --++ (-1.5,0)node[left]{9} --++(1.75,0);
\draw (node9) --++ (0,3) coordinate(node2);
\path (node8) --++ (2,8.0)coordinate(node1);
\path (node8) --++ (2,1.0)coordinate(node10);
\draw (node10) --++ (0.5,0) --++ (0,7.0);
\draw (node10) --++ (-1.75,0)node[left]{10} --++(2.5,0);
\draw (node10) --++ (-1.5,0) --++ (0,2);
\path (node10) --++ (-0.5,0)coordinate(node10l);
\path (node10l) --++ (-0.5,0) coordinate(node10line);
\path (node10l) --++ (0.25,0) coordinate(node10HVDC);
\draw[red] (node10line) --++ (0,5);
\draw[-{Latex[length=0.25cm]}](node10l) --++(0,-0.75)node[below]{};
\path (node9) --++ (-0.5,0)coordinate(node9l);
\path (node1) --++ (-0.15,0) coordinate(node1l);
\draw[-{Latex[length=0.25cm]}](node1l) --++(0,-0.75)node[below]{};
\draw[-{Latex[length=0.25cm]}](node9l) --++(0,-0.75)node[below]{};
\draw (node1) --++ (-1,0)node[left]{1} --++(2,0);
\draw (node2) --++ (-2,0)node[left]{2} --++(3.5,0);
\draw[blue!80] (node10HVDC) --++ (0,5);
\path (node10HVDC) --++ (-0.3750,0.25)coordinate(node10HVDCu);
\path (node10HVDC) --++ (-0.3750,3.75)coordinate(node10HVDCu2);
\draw[blue!80,fill=white] (node10HVDCu) --++ (0.75,0) --++ (0,0.75) --++ (-0.75,0) --++ (0,-0.75);
\draw[blue!80] (node10HVDCu) --++ (0.75,0.75);
\path[blue] (node10HVDCu) --++ (0.14,0.77) coordinate[label = {above: C2 }] (ymax);
\path[blue!80]  (node10HVDCu) --++ (0.2,0.7) coordinate[label = {below: \Huge{\textasciitilde} }] (ymax);  
\path[blue!80] (node10HVDCu) --++ (0.55,0) coordinate[label = {above: = }] (ymax);
\draw[blue!80,fill=white] (node10HVDCu2) --++ (0.75,0) --++ (0,0.75) --++ (-0.75,0) --++ (0,-0.75);
\draw[blue!80] (node10HVDCu2) --++ (0.75,0.75);
\path[blue!80]  (node10HVDCu2) --++ (0.2,0.7) coordinate[label = {below: \Huge{\textasciitilde} }] (ymax);  
\path[blue!80] (node10HVDCu2) --++ (0.55,0) coordinate[label = {above: = }] (ymax);
\path[blue] (node10HVDCu2) --++ (0.75,0) coordinate[label = {below: C1 }] (ymax);
\draw (node8) --++ (0.25,0) coordinate(node8g);
\path (node8g) --++ (0.25,-0.75) node[right]{G5};
\draw (node8g) --++ (0,-0.5) node[below,sin v source]{};
\path (node2) --++ (-0.5,0) coordinate (node2g);
\draw (node2g) --++ (0,-0.5) node[below,sin v source]{};
\path (node2g) --++ (-0.25,-0.75) node[left]{G1};
\draw (node7) --++ (1.25,0.0) coordinate(node7g);
\draw (node7g) --++ (0,-0.5) node[below,sin v source]{};
\path (node7g) --++ (0.25,-0.75) node[right]{G4};
\path (node7) --++ (0.5,0) coordinate(L7);
\draw [-{Latex[length=0.25cm]}] (L7) --++(0,-0.75)node[below]{};
\path (node4) --++ (-0.5,0) coordinate(L4);
\draw [-{Latex[length=0.25cm]}] (L4) --++(0,-0.75)node[below]{};
\path (node10) --++ (0.3,0) coordinate(node10w);
\draw (node10w) --++ (0,-0.5) node[below,sin v source]{};
\path (node10w)--++ (0.25,-0.75) node[right]{W1};
\end{tikzpicture}
}
\vspace{-0.2cm}
\caption{10 bus system with two wind farms located at buses 4 and 10. An HVDC line (marked in blue) replaces the congested AC line (marked in red) between buses 2 and 10.}
\label{10bus}
\vspace{-0.2cm}
\end{figure}
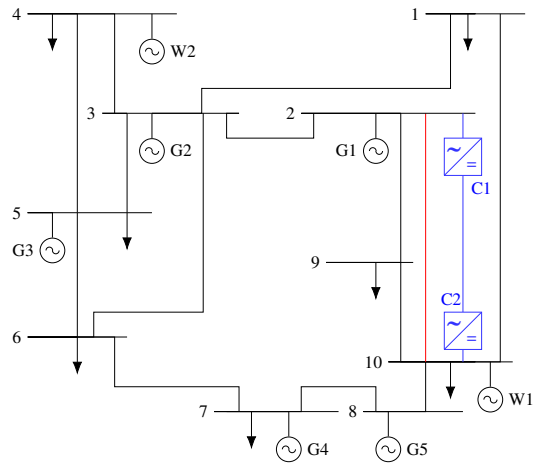
The 10 bus system which is considered in the following first two subsections is shown in Fig.~\ref{10bus}. The grid parameters are provided in \cite{chatzivasileiadis2011flexible}. The generator at bus 3 is selected to be the slack bus. Upper and lower voltage limits of $1.1 \, \text{p.u.}$ and $0.9  \, \text{p.u.}$ are assumed. As we consider the active branch flow limit we set the maximum active branch limit to $80\%$ of the apparent branch flow limit. In this system configuration, the flow of power is from the upper left to the main load units at buses 7 to 10 and the transmission line from bus 2 to bus 10 is congested. Two wind farms are located at buses 10 and 4 with a maximum power of 1.0 \,GW and of 2.5 \,GW, respectively. To compute the covariance matrix $\mathbf{\Sigma}$ of the forecast errors, we use realistic day-ahead  wind  forecast  scenarios  from \cite{pinson2013}.  The  forecasts are based on wind power measurements in the Western Denmark  area  from  15  different  control  zones  collected  by  the Danish  transmission  system  operator  Energinet.  We  select control  zone  7 and  9 at time step 4 to  correspond  to  the  wind  farms at  bus 2 and 10,  respectively. In order to construct the covariance matrix we draw 100 random samples from this data. The forecasted wind infeed is computed as the mean of these 100 samples. Note, for the 10 bus system, due to the small system size, we do not employ the constraint generation method proposed in Section~\ref{Sol_algo} but we directly solve the OPF problem with all uncertainty margins included. For the 39 bus system we employ the constraint generation method to maintain scalability.


  \subsection{Optimization of Generator Participation Factors}
  In this section, for the 10 bus test case, we show the benefit in terms of generation cost of optimizing the generator participation factors $\alpha$ instead of assigning uniform participation factors.  The fixed participation factors are chosen to be $\alpha = [0.2 \,  0.2\, 0.2 \,0.2 \,0.2]$, i.e. each generator equally compensates the deviation in wind power. We compare the performance of an AC-OPF without considering uncertainty, the iterative chance-constrained AC-OPF (CC-AC-OPF) with fixed generator participation factors and the latter (CC-AC-OPF) with optimizing the generator participation factors. \bl{For the 10 bus test case, the overall dispatch cost without considering uncertainty (uncertainty margins set to zero) is $27.14\times10^5 \tfrac{\$}{\text{h}}$, with considering uncertainty and fixed participation factors is $27.69\times10^4 \tfrac{\$}{\text{h}}$ and with optimizing the participation factors is $27.25\times10^4 \tfrac{\$}{\text{h}}$}. As a result, the cost of uncertainty evaluates to $2.03\%$ for fixed participation factors. This can be reduced to $0.39\%$ by optimizing the participation factors. \blue{The number of iterations for fixed $\alpha$ is 5 and for variable $\alpha$ is 6. The average solving time for the AC-OPF iteration is 0.4 seconds for fixed  $\alpha$ and 0.9 seconds for optimizing $\alpha$ as the computational complexity is increased by the including $\alpha$ as optimization variable in the uncertainty margins \eqref{eq:CCACOPF_INEQ2}--\eqref{eq:CCACOPF_INEQ3}.} 
  
  \begin{table}[]
     \centering
          \caption{Empirical constraint violation probability for 10 bus test case without HVDC line} 
     \begin{tabular}{l c c c c}
     \toprule
          Constraint limits on &  $\mathbf{P_G}$ & $\mathbf{Q_G}$ & $\mathbf{V}$ & $\mathbf{P_L}$  \\
     \midrule 
     \multicolumn{5}{c}{In-sample analysis with 10'000 samples (\%)} \\
     \midrule
         AC-OPF (\bl{$\lambda^x = 0$}, w/o uncertainty) & 49.0 & 0.0 & 6.7 & 49.7 \\
         CC-AC-OPF (fixed $\alpha$)  & 5.3 & 0.0 & 2.8 & 5.3 \\
         CC-AC-OPF (opt. $\alpha$) & 4.9 & 0.0 & 2.9 & 4.9 \\
         \midrule
              \multicolumn{5}{c}{Out-of-sample analysis with 10'000 samples (\%)} \\ 
     \midrule 
         AC-OPF (\bl{$\lambda^x = 0$}, w/o uncertainty) & 43.2 & 0.0 & 4.6 & 49.2 \\
         CC-AC-OPF (fixed $\alpha$)  & 5.8 & 0.0 & 3.4 & 6.1 \\
         CC-AC-OPF (opt. $\alpha$) & 5.8 & 0.0 & 3.4 & 5.6 \\
         \bottomrule
         \end{tabular}
     \label{MCA_alpha}
      \vspace{-0.4cm}
 \end{table}
 The results for the Monte Carlo Analysis for in- and out-of-sample testing are shown in Table~\ref{MCA_alpha}. Both in the in- and out-of-sample analyses the AC-OPF without considering uncertainty leads to large empirical violation probabilities for the active generator limits and the active branch flow limits as the response of generators to the wind power deviations is not considered. Voltage violations are observed as well. In case we use the proposed iterative chance-constrained AC-OPF with fixed and optimized generator participation factors we reduce the empirical violation probability both in- and out-of-sample very close to the desired $5\%$. The remaining minor mismatch can be either attributed to a wrong estimation of the mean and covariance in the out-of-sample analysis or to the approximation we make by using the first-order Taylor expansion to linearize the system behaviour around the forecasted operating point. Note that the forecast errors drawn from the realistic forecast data are not Gaussian distributed and the observed violations out-of-sample can therefore be larger. However, we observe that they are still close to the desired $5\%$ indicating good performance of the proposed algorithm.
 
  \begin{figure}
\begin{flushleft}
\begin{footnotesize}
          \begin{tikzpicture}
\begin{axis}[
    ybar=2pt,
     bar width=3pt,
    ylabel={$\mathbf{P_G} / \mathbf{P_G^{\text{max}}} (-)$ },
    symbolic x coords={G1, G2, G3, G4, G5},
    legend style={at={(0.5,-0.15)},
    anchor=north,legend columns=-1},
     width=7.5cm,
     height=3cm,
     xlabel={ (a) -- Normalized generation dispatch},
    ]
\addplot[black,fill] coordinates {(G1, 1.0000) (G2, 0.6993) (G3, 0.9312) (G4, 1.0000) (G5, 0.2583) };
\addplot[blue,fill] coordinates {(G1, 0.9274) (G2, 0.7052) (G3, 0.9534) (G4, 0.8911) (G5, 0.2929) };
\addplot[orange,fill] coordinates {(G1, 1.0000) (G2, 0.6960) (G3, 0.9178) (G4, 1.0000) (G5, 0.2873) };
\end{axis}
\end{tikzpicture} 
     \begin{tikzpicture}
\begin{axis}[
    ybar=2pt,
     bar width=3pt,
    enlargelimits=0.15,
    ylabel={$\lambda^{\mathbf{P_G}}$ (MW) },
    symbolic x coords={G1, G2, G3, G4, G5},
    legend style={at={(0.5,-0.75)},
    anchor=north,legend columns=-1},
     width=7.5cm,
     height=3cm,
    ymin = 0,
    ymax = 300,
     xlabel={ (b) -- Uncertainty margins for active power},
    ]
\addplot[black,fill] coordinates {(G1, 0) (G2, 0) (G3, 0) (G4, 0) (G5, 0) };
\addplot[blue,fill] coordinates {(G1,  87.0884) (G2,   35.3795) (G3,  87.0884) (G4,  87.0884) (G5,  87.0884) };
\addplot[orange,fill] coordinates {(G1, 0.0) (G2,117.3273) (G3, 246.7340) (G4, 0.0) (G5, 56.2219) };
\legend{AC-OPF, CC-AC-OPF (fixed $\alpha$),  CC-AC-OPF (opt. $\alpha$)}
\end{axis}
\end{tikzpicture}
\end{footnotesize}
     \caption{A comparison of (a) normalized generation dispatch and (b) uncertainty margins for active power for AC-OPF without considering uncertainty and the chance-constrained AC-OPF with fixed and optimized generator participation factors. Note that lower active limits of all generators is zero.}
     \label{Alpha_opt}
     \end{flushleft}
      \vspace{-0.4cm}
 \end{figure}
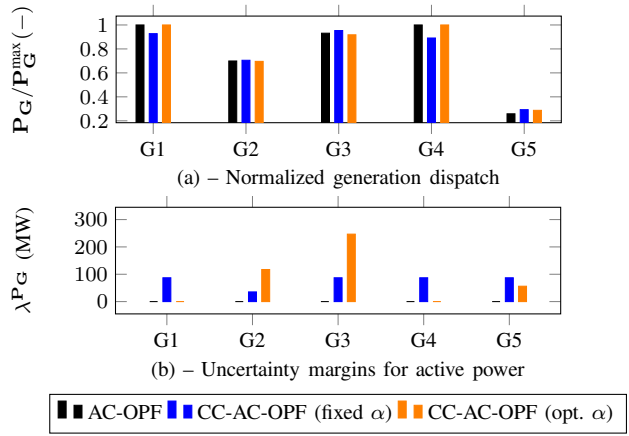
  If we optimize the generator participation factors, we obtain  $\alpha^{\text{opt}} = [0.0 \,\,  0.30 \,\,  0.57 \,\,  0.0 \,\,  0.13]$. In Fig.~\ref{Alpha_opt} we compare the generation dispatch and the uncertainty margins for the three formulations. We can observe that by optimizing the participation factors the generator response is shifted to the generators G2, G3 and G5 with mainly generator G3 compensating the wind power mismatch. The cheap generators G1 and G4 operate at their maximum power output for the forecasted system operating state. This significantly reduces the cost of uncertainty from 2.03\% to 0.39\% while maintaining system reliability.
  \subsection{Including HVDC Line and HVDC Control Policies}
       \begin{table}[!t]
     \centering
          \caption{Empirical constraint violation probability for 10 bus test case with HVDC line} 
     \begin{tabular}{l c c c c c}
     \toprule
          Constraint limits on & $\mathbf{P_G}$ & $\mathbf{Q_G}$ & $\mathbf{V}$ & $\mathbf{P_L}$  & $\mathbf{P_{C}}$ \\
     \midrule 
   \multicolumn{6}{c}{In-sample analysis with 10'000 samples (\%)} \\
     \midrule
         AC-OPF (\bl{$\lambda^x = 0$}, w/o uncertainty) & 50.5 & 0.0 & 45.3 & 12.4 & 0.0 \\
         CC-AC-OPF (fixed $\alpha$ and $\beta$)  & 5.1 & 0.0 & 3.8 & 3.8 & 0.0 \\
         CC-AC-OPF (opt. $\alpha$ and $\beta$) & 0.9 & 0.0 & 3.9 & 3.5 & 0.0 \\
         \blue{CC-AC-OPF (mod.)} & 4.8 & 0.0 & 2.0 & 3.8 & 4.6 \\
         \midrule
              \multicolumn{6}{c}{Out-of-sample analysis with 10'000 samples (\%)} \\
     \midrule 
         AC-OPF (\bl{$\lambda^x = 0$}, w/o uncertainty) & 43.2 & 0.0 &47.8 & 11.5 & 0.0 \\
         CC-AC-OPF (fixed $\alpha$ and $\beta$)  &  5.8& 0.0 & 3.4 & 3.9  & 0.0 \\
         CC-AC-OPF (opt. $\alpha$ and $\beta$) &  0.4 & 0.0 & 3.2 & 3.8 & 0.0 \\
         \blue{CC-AC-OPF (mod.)} & 5.7 & 0.0 & 1.0 & 4.6 & 4.0 \\
         \bottomrule
         \end{tabular}
     \label{MCA_alpha_beta}
      \vspace{-0.4cm}
 \end{table}
 We replace the AC line between buses 2 and 10 in Fig.\ref{10bus} with an HVDC line of $S_C^{\text{nom}}= 4$\,GVA, and investigate the relief of congestion and decrease of the cost of uncertainty.  We assume the converters are of the multi-modular converter (MMC) technology and that the total losses per converter station are approximately $c=1\% $ per HVDC converter according to \cite{jones2013calculation}, and for the active and reactive power capability of the converter the limits are chosen as $m_P = 0.8$, $m_q^{\text{min}} =0.4$, $m_q^{\text{max}} = 0.5$ \cite{imhof2015voltage}. The generation cost for the AC-OPF without considering uncertainty is decreased by 4.3\% \bl{to $25.97\times10^5 \tfrac{\$}{\text{h}}$} due to upgrading the AC to the HVDC line and thereby reducing the congestion level of the system.
 In case we again assume fixed generator participation factors $\alpha = [0.2 \,\, 0.2 \,\, 0.2 \,\, 0.2 \,\, 0.2]$ and HVDC participation factor $\beta = 0$, the cost of uncertainty amounts to 2.2\%. By optimizing both the generator and HVDC participation factors, the cost of uncertainty can be reduced to 0.0\%, i.e. the available HVDC and generator controls are sufficient to absorb the uncertainty associated with the two wind farms without any cost increase. \blue{The number of iterations for both fixed and variable $\alpha$ and $\beta$ is 6. The average solving time for the AC-OPF iteration is 0.4 seconds for fixed $\alpha $ and $\beta$ and is 1.6 seconds for optimizing $\alpha$ and $\beta$, indicating that the computational complexity is further increased by considering $\beta$ as an optimization variable.} 

In Table~\ref{MCA_alpha_beta}, the empirical constraint violation probability for an AC-OPF without considering uncertainty, an iterative CC-AC-OPF with fixed $\alpha$ and $\beta$ and an iterative CC-AC-OPF with optimized $\alpha$ and $\beta$  is shown. We observe again that without considering uncertainty, large violations of the generator active, voltage, and active branch flow limits occur. Both the CC-AC-OPF with fixed and optimized $\alpha$ and $\beta$ achieve a satisfactory performance in- and out-of-sample. For the considered test case, the optimized generator participation factors evaluate to $\alpha = [0.0 \,\, 0.0 \,\, 1.0 \,\, 0.0 \,\, 0.0]$ and the optimized  HVDC participation factor $\beta$ evaluates to 0.1032.

\begin{figure}
    \begin{flushleft}

    \begin{footnotesize}
    	\input{./UM/PG.tex} 
    	\input{./UM/QG.tex}
    	\input{./UM/VG.tex}
    	\input{./UM/PL.tex}
    	\input{./UM/PC.tex}
    	\input{./UM/PF.tex}
    	\end{footnotesize}
    	    \end{flushleft}
    	    \vspace{-0.5cm}
    \caption{\blue{Uncertainty margins $\mathbf{\lambda}$ and participation factors $\alpha$, $\beta$ for each iteration of the chance constrained AC-OPF framework for the 10 bus test system with one HVDC line. The participation factors as optimization variables.}}
    \label{UM}
\end{figure}
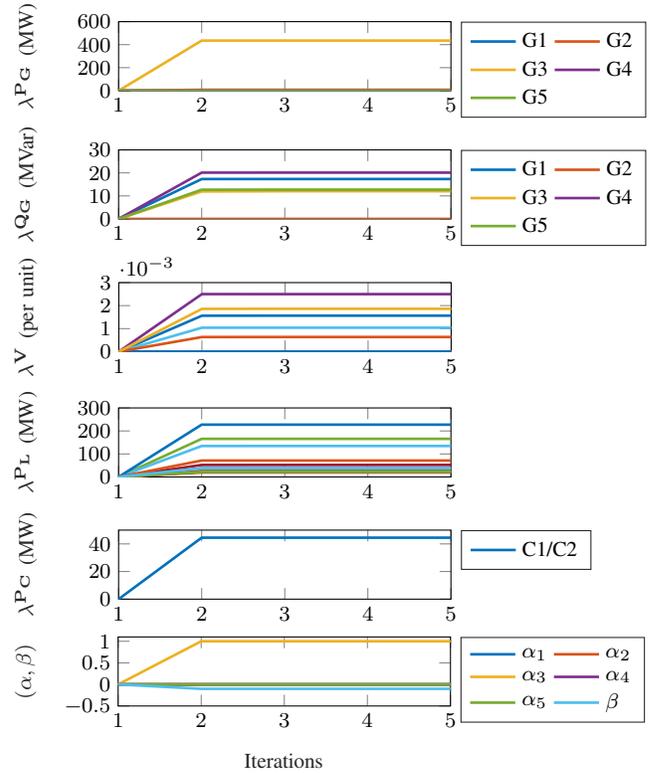
\blue{In Fig.~\ref{UM} the uncertainty margins and participation factors for each iteration of the chance constrained AC-OPF framework are shown for the 10 bus test system with one HVDC line. The participation factors are optimization variables. Note that in the first iteration, the Jacobians are not available. We can observe that after the second iteration the uncertainty margins do not vary significantly showcasing the robustness of the iterative solution framework. The convergence behaviour of the iterative solution algorithm without considering the participation factors as optimization variables is investigated in detail in   \cite{roald2017AnalysisIter}.}

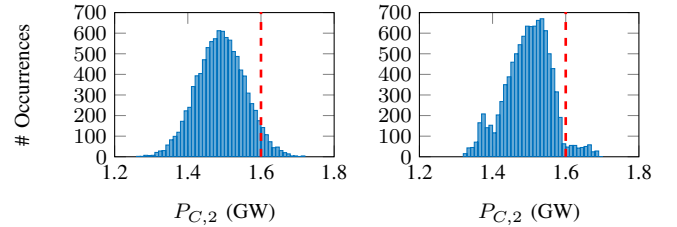
\begin{figure}
\begin{footnotesize}
\input{Hist_In_Sample.tex} \,  \input{Hist_Out_Of_Sample.tex}
\end{footnotesize}
\vspace{-0.4cm}
\caption{\blue{Histograms of the in-sample (left) and out-of-sample (right) Monte Carlo analysis for the active HVDC converter injection $P_{C,2}$ at bus 2 for 10'000 samples. Note, that both in- and out-of-sample the empirical violation probability (4.6\%, 4.0\%) complies with the target value of 5\%. The red dashed line indicates the maximum active power limit of the HVDC converter.}}
\label{Hist}
\end{figure}
\blue{To investigate the ability of the introduced framework to comply with the chance constraints on the active HVDC converter set-points \eqref{eq:UM_HVDC1} -- \eqref{eq:UM_HVDC2}, we consider a modified setup, where the HVDC line capability $S_C^{\text{nom}}$ is reduced to 2 GVA, resulting in congestion on the HVDC line. We assign a fixed participation factor of $\beta = 0.25$ to this HVDC line, and allow for an optimization of the generator participation factors $\alpha$. The resulting empirical violation probability is shown in Table~\ref{MCA_alpha_beta} with the entry CC-AC-OPF (mod.) and achieves satisfactory performance as well. Note, that both in- and out-of-sample the empirical violation probability of the HVDC chance constraints (4.6\%, 4.0\%) complies with the target value of 5\%. This is confirmed in Fig.~\ref{Hist} which shows a histogram of the in- and out-of-sample analysis for the HVDC converter active power injection $P_{C,2}$ at bus $2$.}

\subsection{IEEE 39 bus New England system}
\label{39bus_results}
\blue{In the following, we investigate the performance of our proposed iterative chance constrained AC-OPF algorithm on an IEEE 39 bus New England system with 2 HVDC lines and 2 wind farms. We obtained the system data from the IEEE PES PGLib-OPF v19.01 benchmark library \cite{PGLIB}. We place two farms at buses 4 and 16 with a maximum power of 0.5 \,GW and of 1.0 \,GW, respectively. The maximum wind power injection corresponds to 24.0\% of the total system load. To compute the covariance matrix, and forecast errors, we follow the same procedure as for the 10 bus system. We  select control  zone  7 and  9 at time step 4 to  correspond  to  the  wind  farms at  bus 4 and 16,  respectively. We place two HVDC lines from bus 4 to bus 30 and from bus 16 to bus 38 with $S_C^{\text{nom}} = 500\text{MVA}$, respectively. We assume that only the generators at buses 30, 32, and 36 have a non-zero participation factor $\alpha$. We reduce the line limits to 80\% to obtain a more congested system. For the remaining parameters not specified in \cite{PGLIB} we keep previous assumptions.} 

  \begin{table}[]
     \centering
          \caption{Empirical constraint violation probability for IEEE 39 bus test case with 2 HVDC lines and 2 wind farms} 
     \begin{tabular}{l c c c c c}
     \toprule
          Constraint limits on &  $\mathbf{P_G}$ & $\mathbf{Q_G}$ & $\mathbf{V}$ & $\mathbf{P_L}$ & $\mathbf{P_C}$  \\
     \midrule 
     \multicolumn{6}{c}{In-sample analysis with 10'000 samples (\%)} \\
     \midrule
               AC-OPF (\bl{$\lambda^x = 0$}, w/o uncertainty) & 49.5& 49.3 & 5.3 & 51.3 & 0.0 \\
         CC-AC-OPF (fixed $\alpha$, $\beta$)   &  4.9& 4.2 & 0.0 & 5.5 & 0.0\\
       CC-AC-OPF (opt. $\alpha$, $\beta$) & 4.2 & 3.3 & 0.0 & 5.1 & 4.8 \\
         \midrule
              \multicolumn{6}{c}{Out-of-sample analysis with 10'000 samples (\%)} \\ 
     \midrule 
               AC-OPF (\bl{$\lambda^x = 0$}, w/o uncertainty) & 41.6 & 58.0 & 1.3 & 43.7 & 0.0\\
         CC-AC-OPF (fixed $\alpha$, $\beta$)   & 4.1 & 0.0 & 1.7 & 4.4 &0.0 \\
         CC-AC-OPF (opt. $\alpha$, $\beta$) & 4.1 & 0.0 & 1.2 & 4.3 & 4.2\\
         \bottomrule
         \end{tabular}
     \label{MCA_alpha_beta_39}
      \vspace{-0.4cm}
 \end{table}
\blue{First, we fix the participation factors to be equal in the chance constrained AC-OPF, i.e. \mbox{$\alpha = [\tfrac{1}{3} \tfrac{1}{3} \tfrac{1}{3}]$}, and set the HVDC participation factors to be zero, i.e. \mbox{$\beta = [0\,0]$}. \bl{The overall dispatch cost without considering uncertainty is $11.02\times10^5 \tfrac{\$}{\text{h}}$.} The cost of uncertainty for fixed participation factors evaluates to 1.7\%. If both the generator and HVDC participation factors are optimization variables, for the considered test case, the optimized generator participation factors evaluate to $\alpha = [0.0 \,\, 1.0 \,\, 0.0]$ and the optimized  HVDC participation factors evaluate to $\beta = [0.0 \,\, 0.3540]$. The utilized controllability allows us to reduce the cost of uncertainty to 0.7\%. The average solving time for the iterative AC-OPF is 0.8 seconds with 4 iterations for fixed $\alpha $ and $\beta$ and is 1.6 seconds for optimizing $\alpha$ and $\beta$ with 13 iterations. Note that for this test case, we employ the constraint generation method explained in Section~\ref{Sol_algo}. We observe that only 8 out of the 146 possible uncertainty margins need to be included, thereby reducing the computational effort significantly. In Table~\ref{MCA_alpha_beta_39}, the empirical constraint violation probability for an AC-OPF without considering uncertainty, the iterative CC-AC-OPF with fixed $\alpha$ and $\beta$ and the iterative CC-AC-OPF with optimized $\alpha$ and $\beta$  is shown. We observe that without considering uncertainty, in this test case, large violations of the generator active and reactive power limits occur. Both the CC-AC-OPF with fixed and optimized $\alpha$ and $\beta$ achieve a satisfactory performance in- and out-of-sample.}

\section{Conclusions}
\label{SVI}
In this work, we propose an AC optimal power flow formulation that a) considers uncertainty in wind power infeed, b) incorporates an HVDC line model and c) allows for an optimization of the generator and HVDC control response to fluctuations in renewable generation. For this purpose, we propose a computationally efficient iterative chance-constrained AC-OPF formulation extending \cite{schmidli2016stochastic, roald_andersson_TPWRS2017}. Using realistic wind forecast data and a Monte Carlo Analysis, for 10 and IEEE 39 bus systems with HVDC lines and wind farms, we show that our proposed chance constrained OPF formulation achieves good in- and out-of-sample performance whereas not considering uncertainty leads to high empirical constraint violation probabilities. In addition, we find that optimizing the participation factors reduces the cost of uncertainty significantly. \bl{Our directions for future work are twofold: First, the presented framework could be extended to take into account interconnected AC and HVDC grids, in particular DC buses with multiple HVDC line connections. Second, data-driven approaches such as \cite{halilbavsic2018data,konstantelos2016implementation} could be incorporated to include stability criteria (e.g. small-signal stability) in the chance-constrained OPF by encoding the feasible space using mixed integer programming.}

\section*{Acknowledgment}
This work  is  supported  by  the  multiDC  project,  funded  by  Innovation  Fund Denmark, Grant Agreement No. 6154-00020B. The authors would like to thank Pierre Pinson for sharing the forecast data.

\bibliographystyle{IEEEtran}
\bibliography{Bib}
\end{document}

%% file: UM/PG.tex
%
%
\definecolor{mycolor1}{rgb}{0.00000,0.44700,0.74100}%
\definecolor{mycolor2}{rgb}{0.85000,0.32500,0.09800}%
\definecolor{mycolor3}{rgb}{0.92900,0.69400,0.12500}%
\definecolor{mycolor4}{rgb}{0.49400,0.18400,0.55600}%
\definecolor{mycolor5}{rgb}{0.46600,0.67400,0.18800}%
\begin{tikzpicture}

\begin{axis}[%
width=6cm,
height=2.5cm,
xmin=1,
xmax=5,
xlabel style={font=\color{white!15!black}},
ymin=0,
ymax=600,
ylabel style={font=\color{white!15!black}},
ylabel={$\lambda^{\mathbf{P_G}}$ (MW)},
axis background/.style={fill=white},
legend style={legend cell align=left, align=left, draw=white!15!black,legend pos=outer north east},
legend columns=2, 
]
\addplot [color=mycolor1,line width = 1pt]
  table[row sep=crcr]{%
1	0\\
2	0\\
3	0\\
4	0\\
5	0\\
};
\addlegendentry{G1}

\addplot [color=mycolor2,line width = 1pt]
  table[row sep=crcr]{%
1	0\\
2	7.77741684432813\\
3	7.79302997439108\\
4	7.79304115228137\\
5	7.79304095330713\\
};
\addlegendentry{G2}

\addplot [color=mycolor3,line width = 1pt]
  table[row sep=crcr]{%
1	0\\
2	435.442101571922\\
3	435.442101571922\\
4	435.442101571922\\
5	435.442101571922\\
};
\addlegendentry{G3}

\addplot [color=mycolor4,line width = 1pt]
  table[row sep=crcr]{%
1	0\\
2	0\\
3	0\\
4	0\\
5	0\\
};
\addlegendentry{G4}

\addplot [color=mycolor5,line width = 1pt]
  table[row sep=crcr]{%
1	0\\
2	0\\
3	0\\
4	0\\
5	0\\
};
\addlegendentry{G5}

\end{axis}
\end{tikzpicture}%

%% file: UM/QG.tex
%
%
\definecolor{mycolor1}{rgb}{0.00000,0.44700,0.74100}%
\definecolor{mycolor2}{rgb}{0.85000,0.32500,0.09800}%
\definecolor{mycolor3}{rgb}{0.92900,0.69400,0.12500}%
\definecolor{mycolor4}{rgb}{0.49400,0.18400,0.55600}%
\definecolor{mycolor5}{rgb}{0.46600,0.67400,0.18800}%
\begin{tikzpicture}

\begin{axis}[%
width=6cm,
height=2.5cm,
xmin=1,
xmax=5,
xlabel style={font=\color{white!15!black}},
ymin=0,
ymax=30,
ylabel style={font=\color{white!15!black}},
ylabel={$\lambda^{\mathbf{Q_G}}$ (MVar)},
axis background/.style={fill=white},
legend style={legend cell align=left, align=left, draw=white!15!black,legend pos=outer north east},
legend columns=2, 
]
\addplot [color=mycolor1,line width = 1pt]
  table[row sep=crcr]{%
1	0\\
2	17.3305567585936\\
3	17.3386439144744\\
4	17.3386588634549\\
5	17.3386579538677\\
};
\addlegendentry{G1}

\addplot [color=mycolor2,line width = 1pt]
  table[row sep=crcr]{%
1	0\\
2	0\\
3	0\\
4	0\\
5	0\\
};
\addlegendentry{G2}

\addplot [color=mycolor3,line width = 1pt]
  table[row sep=crcr]{%
1	0\\
2	11.9257461969502\\
3	12.0355494681781\\
4	12.0357262267982\\
5	12.0357245690123\\
};
\addlegendentry{G3}

\addplot [color=mycolor4,line width = 1pt]
  table[row sep=crcr]{%
1	0\\
2	20.1131239715562\\
3	20.1268499993822\\
4	20.1268337223115\\
5	20.1268348733167\\
};
\addlegendentry{G4}

\addplot [color=mycolor5,line width = 1pt]
  table[row sep=crcr]{%
1	0\\
2	12.6976534325914\\
3	12.7261021038729\\
4	12.726149487782\\
5	12.7261486565631\\
};
\addlegendentry{G5}

\end{axis}
\end{tikzpicture}%

%% file: UM/VG.tex
%
%
\definecolor{mycolor1}{rgb}{0.00000,0.44700,0.74100}%
\definecolor{mycolor2}{rgb}{0.85000,0.32500,0.09800}%
\definecolor{mycolor3}{rgb}{0.92900,0.69400,0.12500}%
\definecolor{mycolor4}{rgb}{0.49400,0.18400,0.55600}%
\definecolor{mycolor5}{rgb}{0.46600,0.67400,0.18800}%
\definecolor{mycolor6}{rgb}{0.30100,0.74500,0.93300}%
\definecolor{mycolor7}{rgb}{0.63500,0.07800,0.18400}%
\begin{tikzpicture}

\begin{axis}[%
width=6cm,
height=2.5cm,
xmin=1,
xmax=5,
xlabel style={font=\color{white!15!black}},
ymin=0,
ymax=0.003,
ylabel style={font=\color{white!15!black}},
ylabel={$\lambda^{\mathbf{V}}$ (per unit)},
axis background/.style={fill=white},
legend style={legend cell align=left, align=left, draw=white!15!black,legend pos=outer north east},
legend columns=2, 
]
\addplot [color=mycolor1,line width = 1pt, forget plot]
  table[row sep=crcr]{%
1	0\\
2	0.00156124270415088\\
3	0.00156384663620941\\
4	0.00156385076623162\\
5	0.00156385074462871\\
};
\addplot [color=mycolor2,line width = 1pt, forget plot]
  table[row sep=crcr]{%
1	0\\
2	0\\
3	0\\
4	0\\
5	0\\
};
\addplot [color=mycolor3,line width = 1pt, forget plot]
  table[row sep=crcr]{%
1	0\\
2	0\\
3	0\\
4	0\\
5	0\\
};
\addplot [color=mycolor4,line width = 1pt, forget plot]
  table[row sep=crcr]{%
1	0\\
2	0.00250078317594537\\
3	0.00249923534940891\\
4	0.00249923251882591\\
5	0.00249923253442761\\
};
\addplot [color=mycolor5,line width = 1pt, forget plot]
  table[row sep=crcr]{%
1	0\\
2	0\\
3	0\\
4	0\\
5	0\\
};
\addplot [color=mycolor6,line width = 1pt, forget plot]
  table[row sep=crcr]{%
1	0\\
2	0.00103943803390891\\
3	0.00103942445708826\\
4	0.00103941919040735\\
5	0.00103941924796519\\
};
\addplot [color=mycolor7,line width = 1pt, forget plot]
  table[row sep=crcr]{%
1	0\\
2	0\\
3	0\\
4	0\\
5	0\\
};
\addplot [color=mycolor1,line width = 1pt, forget plot]
  table[row sep=crcr]{%
1	0\\
2	0\\
3	0\\
4	0\\
5	0\\
};
\addplot [color=mycolor2,line width = 1pt, forget plot]
  table[row sep=crcr]{%
1	0\\
2	0.000634149821352424\\
3	0.000635676544214243\\
4	0.00063567949671868\\
5	0.00063567940045877\\
};
\addplot [color=mycolor3,line width = 1pt, forget plot]
  table[row sep=crcr]{%
1	0\\
2	0.00185817708575732\\
3	0.00186218161713287\\
4	0.00186218958368235\\
5	0.00186218958630328\\
};
\end{axis}
\end{tikzpicture}%

%% file: UM/PL.tex
%
%
\definecolor{mycolor1}{rgb}{0.00000,0.44700,0.74100}%
\definecolor{mycolor2}{rgb}{0.85000,0.32500,0.09800}%
\definecolor{mycolor3}{rgb}{0.92900,0.69400,0.12500}%
\definecolor{mycolor4}{rgb}{0.49400,0.18400,0.55600}%
\definecolor{mycolor5}{rgb}{0.46600,0.67400,0.18800}%
\definecolor{mycolor6}{rgb}{0.30100,0.74500,0.93300}%
\definecolor{mycolor7}{rgb}{0.63500,0.07800,0.18400}%
\begin{tikzpicture}

\begin{axis}[%
width=6cm,
height=2.5cm,
xmin=1,
xmax=5,
xlabel style={font=\color{white!15!black}},
ymin=0,
ymax=300,
ylabel style={font=\color{white!15!black}},
ylabel={$\lambda^{\mathbf{P_L}}$ (MW)},
axis background/.style={fill=white},
legend style={legend cell align=left, align=left, draw=white!15!black,legend pos=outer north east},
legend columns=2, 
]
\addplot [color=mycolor1, forget plot,line width = 1pt]
  table[row sep=crcr]{%
1	0\\
2	26.843302478911\\
3	26.8411916884989\\
4	26.84117630335\\
5	26.8411758976791\\
};
\addplot [color=mycolor2, forget plot,line width = 1pt]
  table[row sep=crcr]{%
1	0\\
2	19.4504907273935\\
3	19.4346163758179\\
4	19.4345814076283\\
5	19.4345811517106\\
};
\addplot [color=mycolor3, forget plot,line width = 1pt]
  table[row sep=crcr]{%
1	0\\
2	36.9796325900964\\
3	36.9885200690822\\
4	36.98859006863\\
5	36.9885851080912\\
};
\addplot [color=mycolor4, forget plot,line width = 1pt]
  table[row sep=crcr]{%
1	0\\
2	22.3618352665472\\
3	22.3535836836185\\
4	22.3535695212367\\
5	22.3535661643295\\
};
\addplot [color=mycolor5, forget plot,line width = 1pt]
  table[row sep=crcr]{%
1	0\\
2	165.633345986725\\
3	165.662375112946\\
4	165.662112695998\\
5	165.662112648223\\
};
\addplot [color=mycolor6, forget plot,line width = 1pt]
  table[row sep=crcr]{%
1	0\\
2	134.562285819199\\
3	134.595419803473\\
4	134.595285446161\\
5	134.595284699196\\
};
\addplot [color=mycolor7, forget plot,line width = 1pt]
  table[row sep=crcr]{%
1	0\\
2	52.2469735434445\\
3	52.2680142275587\\
4	52.2679846379708\\
5	52.267983301171\\
};
\addplot [color=mycolor1, forget plot,line width = 1pt]
  table[row sep=crcr]{%
1	0\\
2	227.378931705373\\
3	227.336232727048\\
4	227.336492414994\\
5	227.336492590143\\
};
\addplot [color=mycolor2, forget plot,line width = 1pt]
  table[row sep=crcr]{%
1	0\\
2	71.3416636180535\\
3	71.3495227772336\\
4	71.349422517976\\
5	71.3494230429092\\
};
\addplot [color=mycolor3, forget plot,line width = 1pt]
  table[row sep=crcr]{%
1	0\\
2	24.1095970926226\\
3	24.1041555846765\\
4	24.1041220058725\\
5	24.1041233089652\\
};
\addplot [color=mycolor4, forget plot,line width = 1pt]
  table[row sep=crcr]{%
1	0\\
2	23.86929364596\\
3	23.8632593122953\\
4	23.8632233882573\\
5	23.8632246743683\\
};
\addplot [color=mycolor5, forget plot,line width = 1pt]
  table[row sep=crcr]{%
1	0\\
2	22.3618352665472\\
3	22.3535836836185\\
4	22.3535695212367\\
5	22.3535661643295\\
};
\addplot [color=mycolor6, forget plot,line width = 1pt]
  table[row sep=crcr]{%
1	0\\
2	39.0757685709259\\
3	39.0922795783635\\
4	39.09230793029\\
5	39.09230881059\\
};
\end{axis}
\end{tikzpicture}%

%% file: UM/PC.tex
%
%
\definecolor{mycolor1}{rgb}{0.00000,0.44700,0.74100}%
\begin{tikzpicture}

\begin{axis}[%
width=6cm,
height=2.5cm,
xmin=1,
xmax=5,
xlabel style={font=\color{white!15!black}},
ymin=0,
ymax=50,
ylabel style={font=\color{white!15!black}},
ylabel={$\lambda^{\mathbf{P_C}}$ (MW)},
axis background/.style={fill=white},
legend style={legend cell align=left, align=left, draw=white!15!black,legend pos=outer north east},
legend columns=2, 
]
\addplot [color=mycolor1, line width = 1pt]
  table[row sep=crcr]{%
1	0\\
2	44.5207927915878\\
3	44.5138220499024\\
4	44.5138945803349\\
5	44.5138835757014\\
};
\addlegendentry{C1/C2}
\end{axis}
\end{tikzpicture}%

%% file: UM/PF.tex
%
%
\definecolor{mycolor1}{rgb}{0.00000,0.44700,0.74100}%
\definecolor{mycolor2}{rgb}{0.85000,0.32500,0.09800}%
\definecolor{mycolor3}{rgb}{0.92900,0.69400,0.12500}%
\definecolor{mycolor4}{rgb}{0.49400,0.18400,0.55600}%
\definecolor{mycolor5}{rgb}{0.46600,0.67400,0.18800}%
\definecolor{mycolor6}{rgb}{0.30100,0.74500,0.93300}%
\begin{tikzpicture}

\begin{axis}[%
width=6cm,
height=2.5cm,
xmin=1,
xmax=5,
xlabel style={font=\color{white!15!black}},
xlabel={Iterations},
ymin=-0.5,
ymax=1.1,
ylabel style={font=\color{white!15!black}},
ylabel={$(\alpha,\beta)$ },
axis background/.style={fill=white},
legend style={legend cell align=left, align=left, draw=white!15!black,legend pos=outer north east},
legend columns=2, 
]
\addplot [color=mycolor1,line width = 1pt]
  table[row sep=crcr]{%
1	0\\
2	0\\
3	0\\
4	0\\
5	0\\
};
\addlegendentry{$\alpha_1$}

\addplot [color=mycolor2,line width = 1pt]
  table[row sep=crcr]{%
1	0\\
2	2.137398944813e-08\\
3	2.13955400822061e-08\\
4	2.14509287519295e-08\\
5	2.13578486018246e-08\\
};
\addlegendentry{$\alpha_2$}

\addplot [color=mycolor3,line width = 1pt]
  table[row sep=crcr]{%
1	0\\
2	1\\
3	1\\
4	1\\
5	1\\
};
\addlegendentry{$\alpha_3$}

\addplot [color=mycolor4,line width = 1pt]
  table[row sep=crcr]{%
1	0\\
2	0\\
3	0\\
4	0\\
5	0\\
};
\addlegendentry{$\alpha_4$}

\addplot [color=mycolor5,line width = 1pt]
  table[row sep=crcr]{%
1	0\\
2	0\\
3	0\\
4	0\\
5	0\\
};
\addlegendentry{$\alpha_5$}

\addplot [color=mycolor6,line width = 1pt]
  table[row sep=crcr]{%
1	-0\\
2	-0.10224273819842\\
3	-0.102226729774659\\
4	-0.102226896341999\\
5	-0.102226871069675\\
};
\addlegendentry{$\beta$}

\end{axis}
\end{tikzpicture}%

%% file: Hist_In_Sample.tex
%
%
\definecolor{mycolor1}{rgb}{0.00000,0.44700,0.74100}%
\begin{tikzpicture}

\begin{axis}[%
width=4.5cm,
height=3.5cm,
xmin=1200,
xmax=1800,
ymin=0,
ymax=700,
axis background/.style={fill=white},
xlabel={$P_{C,2}$ (GW)},
ylabel={\# Occurrences},
xtick={1200,1400,1600,1800},
ytick={0,100,200,300,400,500,600,700},
xticklabels ={1.2,1.4,1.6,1.8}
]
\addplot[ybar interval, fill=mycolor1, fill opacity=0.6, draw=mycolor1, area legend] table[row sep=crcr] {%
x	y\\
1260	2\\
1270	2\\
1280	7\\
1290	6\\
1300	7\\
1310	21\\
1320	28\\
1330	30\\
1340	57\\
1350	82\\
1360	99\\
1370	119\\
1380	172\\
1390	223\\
1400	240\\
1410	341\\
1420	393\\
1430	404\\
1440	471\\
1450	530\\
1460	559\\
1470	573\\
1480	612\\
1490	609\\
1500	579\\
1510	566\\
1520	511\\
1530	487\\
1540	421\\
1550	392\\
1560	309\\
1570	258\\
1580	225\\
1590	175\\
1600	142\\
1610	107\\
1620	73\\
1630	44\\
1640	47\\
1650	30\\
1660	20\\
1670	14\\
1680	4\\
1690	4\\
1700	1\\
1710	4\\
1720	4\\
};
\addplot[draw=red,dashed,line width = 1pt]  table[row sep=crcr]{
1600 0 \\
1600.001 800 \\
};
\end{axis}
\end{tikzpicture}%

%% file: Hist_Out_Of_Sample.tex
%
%
\definecolor{mycolor1}{rgb}{0.00000,0.44700,0.74100}%
\begin{tikzpicture}

\begin{axis}[%
width=4.5cm,
height=3.5cm,
xmin=1200,
xmax=1800,
ymin=0,
ymax=700,
axis background/.style={fill=white},
xlabel={$P_{C,2}$ (GW)},
xtick={1200,1400,1600,1800},
ytick={0,100,200,300,400,500,600,700},
xticklabels ={1.2,1.4,1.6,1.8}
]
\addplot[ybar interval, fill=mycolor1, fill opacity=0.6, draw=mycolor1, area legend] table[row sep=crcr] {%
x	y\\
1320	15\\
1330	43\\
1340	45\\
1350	71\\
1360	165\\
1370	208\\
1380	141\\
1390	153\\
1400	116\\
1410	203\\
1420	268\\
1430	350\\
1440	388\\
1450	459\\
1460	500\\
1470	544\\
1480	585\\
1490	634\\
1500	632\\
1510	634\\
1520	662\\
1530	670\\
1540	612\\
1550	500\\
1560	428\\
1570	315\\
1580	191\\
1590	63\\
1600	47\\
1610	55\\
1620	55\\
1630	42\\
1640	45\\
1650	50\\
1660	58\\
1670	24\\
1680	28\\
1690	1\\
1700	1\\
};
\addplot[draw=red,dashed,line width = 1pt]  table[row sep=crcr]{
1600 0 \\
1600.001 800 \\
};
\end{axis}
\end{tikzpicture}%